\documentclass[onecolumn,11pt]{ieeetran}

\usepackage{tikz,units,url}

\usepackage{bbold}
\usepackage{ifthen}
\usepgflibrary{arrows}
\usepgflibrary[arrows]
\usetikzlibrary{arrows}
\usetikzlibrary[arrows]

\usepackage[utf8x]{inputenc}
\usepackage{amsmath,amssymb}
\usepackage{mathtools}
\usepackage{graphicx}
\usepackage{multicol}
\usepackage[squaren,thinspace]{SIunits}

\newcommand{\FB}{\mathrm{Fbeta}}

\newcommand{\calR}{\mathcal{R}}

\newcommand{\fbb}{f\ped{bb}}
\newcommand{\Fbb}{F\ped{bb}}

\newcommand{\ones}{\mathbf{1}}

\newcommand{\bin}{\mathrm{Bin}}

\newtheorem{theorem}{Theorem}

 \newtheorem{lemma}{Lemma}

\newtheorem{remark}{Remark}

\newtheorem{definition}{Definition}

\newtheorem{algorithm}{Algorithm}

 \newtheorem{corollary}{Corollary}
\newtheorem{assumption}{Assumption}

\renewcommand{\sc}{\mathrm{Sc}(K)}

\newcommand{\est}{{\hfill $\star$}}
\newcommand{\tran}{^{\top}}

\renewcommand{\d}{\mbox {\rm d}}

\newcommand{\lam}{\lambda}

\newcommand*{\qed}{\hfill\ensuremath{\square}}

\newcommand{\prob}{\mbox{\rm Prob}}

\newcommand{\Q}{\mathbb{Q}}

\newcommand{\Ev}{\mathbb{E}}

\newcommand{\inv}{^{-1}}
\newcommand{\beq}{\begin{equation}}
\newcommand{\eeq}{\end{equation}}
\newcommand{\bea}{\begin{eqnarray}}
\newcommand{\eea}{\end{eqnarray}}
\newcommand{\beas}{\begin{eqnarray*}}
\newcommand{\eeas}{\end{eqnarray*}}
\newcommand{\ba}{\begin{array}}
\newcommand{\ea}{\end{array}}
\newcommand{\bit}{\begin{itemize}}
\newcommand{\eit}{\end{itemize}}
\newcommand{\ben}{\begin{enumerate}}
\newcommand{\een}{\end{enumerate}}

\newcommand{\Real}[1]{ { {\mathbb R}^{#1} } }

\newcommand{\ped}[1]{{_{\mathrm{#1}}}}

\title{On Repetitive Scenario Design}
\author{Giuseppe C. Calafiore\thanks{Giuseppe C. Calafiore, Dipartimento di Automatica e Informatica, Politecnico di Torino, Italy. {\tt\small giuseppe.calafiore@polito.it}} 
}

\begin{document}


\maketitle


\begin{abstract}
Repetitive Scenario Design (RSD) is a randomized approach to robust design based on iterating two phases: a standard scenario design phase
that uses $N$ scenarios (design samples),
 followed by randomized feasibility phase that uses
 $N_o$ test samples on the scenario solution. 
 We give a full and exact probabilistic characterization of the number of iterations required by the RSD approach for returning a solution, as a function of $N$, $N_o$,
 and of the desired levels of probabilistic robustness in the solution. 
 This novel approach 
 broadens the applicability of the scenario technology, since the user 
 is now presented with a clear tradeoff
 between the number $N$ of design samples and the 
 ensuing expected number of repetitions required by the RSD algorithm.
 The plain (one-shot) scenario design becomes just one of the possibilities, sitting
 at one extreme of the tradeoff curve,
 in which  one insists in finding a solution in a single repetition: this comes at the cost of possibly high $N$. Other possibilities along the tradeoff curve
 use lower $N$ values, but possibly require more than one repetition.
\end{abstract}

\begin{keywords}
Scenario design, probabilistic robustness, randomized algorithms, random convex programs.
\end{keywords}

\section{Introduction}
The  purpose of the approach described in this paper is to obtain a {\em probabilistically reliable} solution for some design problem affected by uncertainty.
The concept of ``probabilistic design'' has been discussed extensively in the control community in the last decade, and it is now well accepted as 
a standard tool for tacking difficult robust design problems;
we refer the reader to the  survey paper \cite{CaDaTe:11} and to the book \cite{TeCaDa:13} for many pointers to the related  literature.
The essential elements of a probabilistic design approach are the following ones: 
\ben
\item A {\em spec function}, $f(\theta,q): \Real{n}\times \Q \to \Real{}$, 
which associates a real value to each pair $(\theta,q)$  of a design parameter $\theta\in\Real{n}$ and  uncertainty instance $q\in\Q$, where $\Q\subseteq \Real{n_q}$. Function $f$ represents the design constraints and specifications of the problem and, in particular, we shall say that a design $\theta$ is a {\em robust} design, if
$f(\theta,q)\leq 0$, $\forall q\in\Q$.
In this paper, we make the standing assumption that $f$ is convex.

\item A probability measure $\prob$ defined on $\Q$, which describes the probability distribution of the uncertainty. 
\een
Equipped with these two essential elements, for given $\epsilon \in (0,1)$, and given design vector
$\theta$, we are in position to define the {\em probability of violation} for the spec function at $\theta$:
\beq
V(\theta) \doteq \prob\{q\in\Q:\, f(\theta,q) > 0\}.
\label{eq:probviol}
\eeq
We say that $\theta$ is an {\em $\epsilon$-probabilistic robust design}, if it holds that
$V(\theta) \leq \epsilon$.
Further, a designer also typically seeks to minimize some cost function of $\theta$ (which can be considered of the linear form 
$c\tran\theta$, without loss of generality; see, e.g., Section~8.3.4.4 in \cite{CalElg:14}), while guaranteeing that
$V(\theta) \leq \epsilon$.
Finding such an $\epsilon$-probabilistic robust design 
amounts to solving a so-called {\em chance-constrained} optimization problem, which
is computationally hard in general, and perhaps harder than
finding a classical deterministic robust design. 
Chance-constrained optimization problems can be solved exactly only in very restrictive cases (e.g., when $f$ is linear,
and $q$ has some specific distribution, such as Normal; see, e.g., \cite{Prekopa:03}). Deterministic convex {\em approximations} of chance-constrained  problems are discussed in
\cite{NemSha:09} for some special classes of problems where $f$ is affine in $q$ and the entries of $q$ are independent.
Also, the {\em sampling average approximation} (SAA) method replaces the 
probability constraint $V(\theta) \leq \epsilon$ with one involving the empirical probability of violation based on
$N$ sampled values of $q$; see, e.g., \cite{PaAhSh:09,LueAhm:08}. 
The optimization problem resulting from SAA, however, remains non-convex and intractable, in general.

\subsection{The standard scenario theory}
While effective approximation schemes for chance-constrained optimization problems remain to date 
hard to tackle numerically, an alternative and efficient 
randomized scheme 
emerged in the last decade
 for finding  $\epsilon$-probabilistic robust designs. 
This technique, which is now 
a well-established technology (see, e.g., the recent surveys \cite{GarCam:13,PetTem:14}) in the area of robust control,
is called ``scenario design,'' and was introduced in \cite{CalCam:06tac}. In scenario design one considers 
$N$ i.i.d.\ random samples of the uncertainty
$
\{q^{(1)},\ldots,q^{(N)}\}\doteq \omega
$,
and builds a scenario {\em random convex program} (RCP):
\bea
 \min_{\theta\in\Theta} & c\tran \theta & \label{eq:rcp}\\
\mbox{s.t.:} & f(\theta,q^{(i)})\leq 0, & i=1,\ldots,N, \nonumber
\eea
where $\Theta$ is some given convex and compact domain, and $c$ is the given objective direction.
An optimal solution $\theta^*$ to this problem, if it exists, is a random variable which depends on the multiextraction $\omega$, i.e., $\theta^* = \theta^*(\omega)$. As a consequence, the violation probability relative to a scenario solution, $V(\theta^*)$, is itself, a priori, a random variable.

Scenario design lies somewhere in between worst-case robust design (where $c\tran \theta$ is minimized subject to
$f(\theta,q)\leq 0$ for all $q\in \Q$) and chance-constrained design (where $c\tran \theta$ is minimized subject to
$V(\theta)\leq \epsilon$). Indeed, the optimal objective value resulting from a scenario design is  lower than
the worst-case optimal objective and it is (with high probability) higher than a suitable chance-constrained
 optimal objective (see, e.g., Section~6 in \cite{Calafiore:SIOPT10}). Moreover,  a fundamental feature of
scenario design is that its optimal solution $\theta^*(\omega)$ is feasible with high probability for the chance-constrained problem. This key result is recalled next for the sake of clarity.
We shall work under the following simplifying assumption, which is routinely made in the literature on scenario design; see \cite{CalCam:06tac,CamGar:08}.

\begin{assumption}
	\label{ass:unique}
With probability (w.p.) one with respect to the multi-extraction $\omega =\{q^{(1)},\ldots,q^{(N)}\}$,
problem (\ref{eq:rcp}) is feasible and it attains a unique optimal solution $\theta^*(\omega)$.
\est
\end{assumption}

Also, we need the following standard definition (see, e.g., Definition~4 in \cite{CalCam:06tac})
\begin{definition}
Let $J^* = c\tran\theta^*$ denote the optimal objective value of problem (\ref{eq:rcp}).
Also, for $j=1,\ldots,N$, define
\bea
 J_j^* \doteq \min_{\theta\in\Theta} & c\tran \theta & \nonumber \label{eq:rcpj}\\
\mbox{s.t.:} & f(\theta,q^{(i)})\leq 0, & i\in \{1,\ldots,N\}\setminus j . \nonumber
\eea
The $i$-th constraint in (\ref{eq:rcp}) is said to be a {\em support constraint} if
$ J_j^* < J^*$. 
\est
\end{definition}

A key fact is that, regardless of the problem structure and of $N$, the number of support constraints for problem
(\ref{eq:rcp}) cannot exceed $n$ (the number of decision variables); see, e.g., Theorem~3 in \cite{CalCam:06tac}.
If an instance of problem (\ref{eq:rcp})  happens to have precisely $n$ support constraints, then the problem instance is said to be {\em fully supported (f.s.)}; see Definition 3 in \cite{CamGar:08}, and Definition 2.5 in \cite{Calafiore:SIOPT10}.
If the instances of problem (\ref{eq:rcp})  are fully supported almost surely with respect to the random extraction
$\omega$ of the $N$ constraints, then we say that problem (\ref{eq:rcp})  is fully supported w.p.\ one.
The following key result holds, see 
 Theorem~1 in \cite{CamGar:08}, and
Corollary~3.4 in \cite{Calafiore:SIOPT10}.

\begin{theorem}
	\label{rcp:thm}
	Let Assumption~\ref{ass:unique} hold. Then, for given $\epsilon \in [0,1]$ and $N\geq n$, it holds that
	\bea
	F_V(\epsilon) &\doteq & \prob^N \{\omega: \, V(\theta^*(\omega)) \leq \epsilon \} \label{eq:FV} \\
	&\geq & 
	\sum_{i=n}^{N}\binom{N}{i} \epsilon^i (1-\epsilon)^{N-i} \label{eq:scenariobound} \\
	&\doteq  & 1-\beta_\epsilon(N).  \nonumber
	\eea
	Moreover, the bound (\ref{eq:scenariobound}) is tight, since
	it holds with equality for the class of problems of the form (\ref{eq:rcp})
	that are  fully supported with probability  one.
	\est
\end{theorem}

A remarkable feature of the result in (\ref{eq:scenariobound})  is that it holds irrespective of the probability distribution
assumed on $q$, and that it depends on the problem structure only through the dimension parameter $n$.

\subsection{Scenario problems and Bernoulli trials}
For given $\epsilon\in[0, 1]$ and $N\geq n$, let us consider the following Bernoulli variable associated to problem (\ref{eq:rcp}):
\[
z = z(\omega) = \left\{\ba{ll} 1, & \mbox{if } V(\theta^*(\omega)) \leq \epsilon \\
0, &\mbox{otherwise.} \ea \right.
\]
By the definition in eq.\  (\ref{eq:FV}), the event $z=1$ happens w.p.\  $F_V(\epsilon)$.
One interpretation of eq.\ (\ref{eq:scenariobound})
is thus that each time we solve a scenario problem (\ref{eq:rcp}) we have an a priori probability
$ \geq 1-\beta_\epsilon(N)$ 
of realizing a ``successful design,'' that is of finding a solution $\theta^*$ which is an $\epsilon$-probabilistic robust design, and a probability $\leq \beta_\epsilon(N)$
of realizing a ``failure,'' that is of finding a solution $\theta^*$ which is not $\epsilon$-probabilistic robust.

In the classical scenario theory it is usually prescribed to choose $N$ so to make $\beta_\epsilon(N)$ very small (values as low as $10^{-12}$ are common). This guarantees that the event
$\{ V(\theta^*(\omega)) \leq \epsilon \} $ will happen with ``practical certainty.''
In other words, in such regime, the scenario problem will return an $\epsilon$-probabilistic robust
solution with practical certainty. Moreover, a key feature of scenario theory is that
such high level of confidence can be reached at a  relatively ``cheap'' computational price.
Indeed, considering the condition
 $\beta_\epsilon(N) \leq \beta$ for some given desired probability level $\beta\in(0,1)$,
 and using some fairly standard techniques for bounding the Binomial tail 
 (see, e.g., Corollary~5.1 in \cite{Calafiore:SIOPT10} for the details), one can prove that
 the condition is satisfied for\footnote{
Notice that the expression in (\ref{eq:Nbound}) may be conservative; the exact minimal value of $N$
 can be easily found numerically by searching for the least integer $N$ such that
$ \sum_{i=n}^{N}\binom{N}{i} \epsilon^i (1-\epsilon)^{N-i} \geq 1-\beta$.}
\beq
N \geq \frac{2}{\epsilon}\left(\ln \beta\inv + n - 1 \right).
\label{eq:Nbound}
\eeq
Since $\beta\inv$ appears in the above bound under a logarithm, we indeed see that  $N$ grows gracefully with
the required certainty level $\beta\inv$.
However, there are  cases in which the number $N$ of constraints 
prescribed by (\ref{eq:Nbound}) for reaching the desired confidence levels is just too high
for practical numerical solution. Convex optimization solvers are certainly efficient, 
but there are practical limits on the number of constraints they can deal with; these limits depend on the actual type of convex problem (say, a linear program (LP), or a semidefinite program (SDP)) one deals with.
A critical situation is, for instance, when problem 
(\ref{eq:rcp}) is a semidefinite program (formally, $f$ can be taken as the maximum eigenvalue function of the matrices describing the linear inequality constraints): dealing with an SDP with many thousands of LMI constraints can pose serious practical issues.

\subsection{Contribution}
In this paper we  discuss how a variation of the scenario approach can be used for obtaining 
an $\epsilon$-probabilistic robust
solution with high confidence, using ``small'' values of $N$. More precisely, we are interested in using  scenario optimization in a regime of $N$ for which the right-hand side of eq.~(\ref{eq:scenariobound}) is {\em not} close to one. We shall do so by
solving repeatedly instances of the scenario problem, and checking the result via
a suitable ``violation oracle.''
This novel approach, named {\em repeated scenario design} (RSD),  is discussed in Section~\ref{sec:repetitive}, which contains
all the relevant results. Section~\ref{sec:examples} describes two numerical examples of  robust control design where
the proposed approach is applied. For improving readability, technical proofs are reported in the Appendix.

\subsection{Notation and preliminaries}
We shall make intensive use of the beta and related probability distributions. Some definitions and standard facts are recalled next.
\label{sec:betabin}
We denote by $\mbox{beta}(\alpha,\beta)$ the beta density function with parameters $\alpha>0$, $\beta>0$:
\[
\mbox{beta}(\alpha,\beta; t) \doteq \frac{1}{B(\alpha,\beta)}t^{\alpha -1}(1-t)^{\beta -1},\quad
t\in[0,1],
\]
where
$
 B(\alpha,\beta) \doteq \frac{\Gamma(\alpha)\Gamma(\beta)}{\Gamma(\alpha+\beta)}
$,
and $\Gamma$ is the Gamma function (for $\alpha,\beta$ integers, it holds that 
$B(\alpha,\beta)\inv = \alpha \binom{\alpha+\beta-1}{\beta-1}$).
Also, we denote by
$\FB (\alpha,\beta)$ the cumulative distribution function of the $\mbox{beta}(\alpha,\beta)$  density:
\beas
\FB (\alpha,\beta; t) &\doteq &  \int_{0}^t \mbox{beta}(\alpha,\beta; \vartheta) \d \vartheta,\quad
t\in[0,1] .
\eeas
$\FB (\alpha,\beta; t)$
is the regularized incomplete beta function, and a standard result establishes that, for $\alpha,\beta$ integers, it holds that
\[
\FB (\alpha,\beta; t) = \sum_{i=\alpha}^{\alpha+\beta-1}\binom{\alpha+\beta-1}{i} t^i (1-t)^{\alpha+\beta-1-i}.
\label{eq:Fbetadefint}
\]
The number $x$ of successes in $d$ independent Bernoulli trials each having success probability
$p$ is a random variable with Binomial distribution (which we denote by $\bin (d,p)$); its  cumulative distribution
is given by
\bea
\prob \{x \leq z\} &\!\!\!=\!\!\!  & \prob \{x \leq \lfloor z \rfloor\} = \sum_{i=0}^{\lfloor z \rfloor}\binom{d}{i} t^i (1-t)^{d-i} \nonumber \\
& =& 1-\sum_{i=\lfloor z \rfloor+1}^{d}\binom{d}{i} t^i (1-t)^{d-i} \nonumber \\
 &
= & 1-\FB (\lfloor z \rfloor + 1,d-\lfloor z \rfloor ; t)   \nonumber\\
&\leq &  1-\FB ( z  + 1,d - z  ; t) \label{eq:binomial} \\
&=&  \FB (d-z,z+1; 1-t),  \nonumber
\eea
where $\lfloor z \rfloor$ denotes the largest integer no larger than $z$.
The number $x$ of successes in $d$ binary trials, where each trial has success probability
$p$, and $p$ is itself a random variable with
$\mbox{beta}(\alpha,\beta)$ distribution, 
 is a random variable with so-called beta-Binomial distribution:
for $i=0,1,\ldots,d$,
\beq
\fbb (d,\alpha,\beta; i) \doteq  \binom{d}{i} \frac{B(i+\alpha,d-i+\beta)}{B(\alpha,\beta)}.
\label{eq:betabinomial}
\eeq
The cumulative distribution of a beta-Binomial random variable is given by (see, e.g., \cite{LeeSab:87,Weisstein})
\beas
 \lefteqn{ \Fbb(d,\alpha,\beta; z) \doteq  \prob \{x \leq z\}  =
\sum_{i=0}^{\lfloor z \rfloor} \fbb (d,\alpha,\beta; i) } \nonumber \\
&& = 1- \frac{1}{d+1}\frac{B(\beta+d-z-1,\alpha+z+1)}{B(\alpha,\beta)B(d-z,z+2)} 
 F_d(a,b;z), \label{eq:bbinom:hypergeom}
\eeas
where $ F_d(a,b;z)$ is the generalized hypergeometric function
\beas
\ _3F_2(1,\alpha+z+1,-n+z+1; z+2,-\beta-n+z+2; 1).
\eeas
%

\section{Repetitive scenario design}
\label{sec:repetitive}
This section develops the main idea of this paper. By {\em repetitive scenario design} (RSD)
we here mean an iterative computational approach in which, at each iteration $k$, the scenario problem (\ref{eq:rcp}) is solved and then the ensuing solution $\theta^*_k$ is checked by a violation oracle (either deterministic, or randomized, as illustrated next). If the oracle returns {\tt false}, another iteration is performed; if instead  the oracle returns {\tt true}, the algorithm is terminated and the current solution
$\theta^*_k$ is returned.

In the RSD the user selects a desired probabilistic feasibility level $\epsilon\in (0,1)$, and
a number $N\geq n$ of scenarios to be used in (\ref{eq:rcp}).
We have from Theorem~\ref{rcp:thm} that,
at any iteration $k$, it holds that
\beq
\prob^N\{\omega^{(k)}: V(\theta^*_k) \leq \epsilon \} = F_V(\epsilon)  \geq 1-\beta_\epsilon(N) ,
\label{eq:betabound}
\eeq
where $\omega^{(k)}$ denotes the multisample $\{q^{(1)}_k,\ldots,q^{(N)}_k\}$.
In very elementary terms, each iteration of the RSD method can be thought of as a biased ``coin toss,''
where the probability of a success in a toss (that is, getting $\theta_k^*$ such that
$V(\theta^*_k) \leq \epsilon$) is at least $1-\beta_\epsilon(N)$. 
In our setting, this probability need not be too close to one: the simple idea behind the RSD method
is to repeat the coin toss until we obtain a success, where success is detected by the violation oracle. As one may easily argue intuitively,
the probability of obtaining a success at some point in the algorithm is much higher than the probability of obtaining a success in a single toss.
A similar idea has been recently proposed  in  \cite{CDTVW:14}, where the authors  solve repeatedly a ``reduced-size'' scenario problem, followed by a randomized test of feasibility. The approach and the results in \cite{CDTVW:14}, however, are distinctively different from the ones proposed here. In \cite{CDTVW:14}, the scenario problems are solved using a number $N_k$ of scenarios that grows with the iteration count $k$, up to the value $N\ped{plain}$ that corresponds to the plain, one-shot, scenario design. 
The major shortcoming of the approach and analysis in \cite{CDTVW:14} is  that  
the  number of iterations is not bounded a-priori, either in a deterministic or in a probabilistic sense, and no   tradeoff curve is proposed for the choice of $N_k$ in function of the expected running time of the algorithm. As a result,
there is no a-priori   guarantee
that the algorithm does not reach the final iteration, in which $N_k$ equals $N\ped{plain}$, hence
the worst-case complexity  of the algorithm in \cite{CDTVW:14} can be worse than the one of the plain scenario design method, and an actual reduction of the number of design samples is not theoretically guaranteed.

We shall next analyze precisely the probabilistic features of our RSD algorithm in two cases.
In the first case we assume that an ideal exact feasibility oracle
is available for checking the current solution $\theta^*_k$; this case may be unrealistic in general, but serves for providing an insightful preliminary analysis of the RSD approach. In the second case, we analyze the RSD approach when a practically implementable randomized feasibility oracle is used. 

\subsection{Violation oracles}
\label{sec:oracles}
A {\em deterministic $\epsilon$-violation oracle} ($\epsilon$-DVO) is a ``black box'' which, when given
in input a value of the design variable $\theta$, returns as output a flag value which is {\tt true} if
$V(\theta)\leq \epsilon$, and {\tt false} otherwise. 
Such an oracle may {\em not} realizable computationally in practice, since 
computing the probability in (\ref{eq:probviol}) is numerically hard, in general.
For this reason, we next also introduce a {\em randomized $\epsilon'$-violation oracle} ($\epsilon'$-RVO), which is defined by means of the randomized scheme described next.


\vspace{.3cm}

\noindent
{\em $\epsilon'$-RVO}
({\em Randomized $\epsilon'$-violation oracle})
Input data:   integer $N_{o}$, level $\epsilon'\in[0,1]$, and $\theta\in\Real{n}$.
Output data: a logic flag, {\tt true} or {\tt false}.
\ben
\rm
\item Generate $N_o$ i.i.d.\ samples 
$\omega_o \doteq \{q^{(1)},\ldots, q^{({N_o})}\}$, according to $\prob$.

\item For $i=1,\ldots,N_o$, let $v_i =1$ if $f(\theta,q^{(i)})> 0$ and $v_i=0$ otherwise. 

\item If $\sum_i v_i\leq \epsilon' N_o$, return {\tt true}, else
return {\tt false}.
\een
\vspace{.2cm}

The $\epsilon'$-RVO simply evaluates the empirical probability of violation
on $N_o$ samples, and returns {\tt true} if it is below $\epsilon'$, and {\tt false} otherwise.
A similar type of randomized feasibility oracle has been previously introduced in \cite{CalDab:07},
and used in a probabilistic design setting also in \cite{CaDaTe:11}; see also Section~11.1 in \cite{TeCaDa:13}, and the ``validation'' step proposed in \cite{CDTVW:14}.
However, the $\epsilon'$-RVO we propose in this paper is  different from
the one used in the cited references: the latter exits with a  {\tt false} flag as soon as one
infeasible sample is found, whereas the $\epsilon'$-RVO allows up to 
$\lfloor \epsilon' N_o\rfloor$ infeasible samples before exit. Also, the kind of a priori analysis we develop here for the repetitive scenario design based on the $\epsilon'$-RVO is entirely novel.

\subsection{Repetitive scenario design with ideal oracle} 
\label{sec:RSD_DVO}
We consider the following RSD algorithm, where each repetition consists of a plain scenario optimization step, followed by a feasibility check of the ensuing solution, performed by an exact feasibility oracle.

\begin{algorithm}[RSD with $\epsilon$-DVO]
\label{alg:RSD_DVO}
Input data:   integer $N\geq n$, level $\epsilon\in[0,1]$.
Output data: solution $\theta^*$. Initialization: set iteration counter to $k=1$.
\ben
\rm
\item (Scenario step) Generate $N$ i.i.d.\ samples 
$\omega^{(k)} \doteq \{q^{(1)}_k,\ldots, q^{({N})}_k\}$ according to $\prob$, and
solve scenario problem (\ref{eq:rcp}). Let $\theta^*_k$ be the resulting optimal solution.

\item 
($\epsilon$-DVO step)
If $V(\theta^*_k)\leq \epsilon$, then set flag to {\tt true}, else set it to {\tt false}.

\item (Exit condition)
If flag is {\tt true}, then exit and return current solution $\theta^* \gets \theta^*_k$; else 
set $k \gets k+1$ and goto~1.
\een
\est
\end{algorithm}

\noindent
The following theorem holds.

\vspace{.2cm}
\begin{theorem}
	\label{thm:RSD_DVO}
		Let Assumption~\ref{ass:unique} hold. 
		Given $\epsilon \in [0,1]$ and $N\geq n$, 
		define the running time $K$ of Algorithm~\ref{alg:RSD_DVO} as the value of the iteration counter
		$k$ when the algorithm exits.
		Then:
		\ben
		\item The solution $\theta^*$ returned by Algorithm~\ref{alg:RSD_DVO} is an
		$\epsilon$-probabilistic robust design, i.e., $V(\theta^*) \leq \epsilon$.
		\item The expected running time of Algorithm~\ref{alg:RSD_DVO} is $\leq (1-\beta_\epsilon(N))\inv$, and equality holds if the scenario problem is f.s.\ w.p.\ 1.
		\item The running time of Algorithm~\ref{alg:RSD_DVO} is $\leq k$ with 
		probability $\geq 1-\beta_\epsilon(N)^k$, and equality holds if the scenario problem is f.s.\ w.p.\ 1.	
		\een
			\est
\end{theorem}

\noindent
See Section~\ref{app:thm:RSD_DVO} in the Appendix for a proof of Theorem~\ref{thm:RSD_DVO}.

\begin{remark}[Potential and limits of the RSD approach]\rm
	The preliminary results in Theorem~\ref{thm:RSD_DVO} show the potential of the RSD approach.
%
Suppose that $N$ is chosen so that
	$\beta_\epsilon(N)$ is about, say, $0.4$. This means that a plain (i.e., 
	one-shot) scenario approach has only at least a
	0.6 chance of returning a ``good'' solution (i.e., an $\epsilon$-probabilistic robust design: a  $\theta^*$ such that
	$V(\theta^*)\leq \epsilon$).
	However, we see from point 3 of
	Theorem~\ref{thm:RSD_DVO}
	that there is more than  $1-10^{-9}$ probability
	that Algorithm~\ref{alg:RSD_DVO} returns an $\epsilon$-probabilistic robust design
	within 23 iterations. Further, 
the eventual outcome of
	Algorithm~\ref{alg:RSD_DVO} is $\epsilon$-probabilistic robust with probability one, and the expected number of iterations of the RSD algorithm is just $(1-0.4)\inv =1.67$, in the worst case of a f.s.\ problem.
	
	Theorem~\ref{thm:RSD_DVO} also shows a fundamental limit of the RSD approach: 
 we can decrease $N$ (and hence increase $\beta_\epsilon(N)$) with respect to a plain scenario design approach, but we cannot decrease $N$ too much, for otherwise $\beta_\epsilon(N) \to 1$, and
	the expected number of iterations of Algorithm~\ref{alg:RSD_DVO} tends to $\infty$.
	There is thus a fundamental tradeoff between the reduction of  $N$ (which reduces the effort needed for solving the scenario problem) and the increase of the number of iterations of Algorithm~\ref{alg:RSD_DVO}. This tradeoff can be fully captured by plotting
the expected running time bound	$(1-\beta_\epsilon(N))\inv$ versus the number $N$ of scenarios.
\end{remark}

\subsection{Repetitive scenario design with randomized oracle}
\label{sec:RSD_RVO}
This section contains the main contribution of this paper, where we
consider a realistically implementable version of the RSD approach,
in which a  randomized oracle is used instead of the ideal deterministic one.

\begin{algorithm}[RSD with $\epsilon'$-RVO]
\label{alg:RSD_RVO}
Input data:   integers $N$, $N_o$, level $\epsilon'\in[0,1]$.
Output data: solution $\theta^*$. Initialization: set iteration counter to $k=1$.
\ben
\rm
\item (Scenario step) Generate $N$ i.i.d.\ samples 
$\omega^{(k)} \doteq \{q^{(1)}_k,\ldots, q^{({N})}_k\}$ according to $\prob$, and
solve scenario problem (\ref{eq:rcp}). Let $\theta^*_k$ be the resulting optimal solution.

\item 
($\epsilon'$-RVO step)
Call the $\epsilon'$-RVO with current $\theta_k^*$ as input, and
 set flag 
 to {\tt true} or {\tt false} according to the output of the $\epsilon'$-RVO.

\item (Exit condition)
If flag is {\tt true}, then exit and return current solution $\theta^* \gets \theta^*_k$; else 
set $k \gets k+1$ and goto 1.
\een
\est
\end{algorithm}

\begin{figure}[htb]
\begin{center}
\includegraphics[width=.75\textwidth]{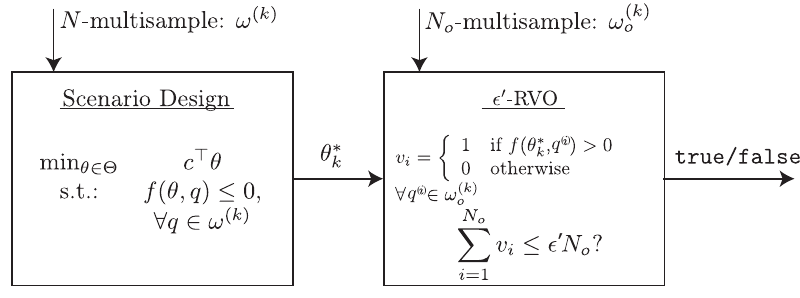}
\caption{Generic stage $k$ of Algorithm~\ref{alg:RSD_RVO}. \label{fig:RSD_stage} }%
\end{center}%
\end{figure}

\noindent
A generic iteration, or stage, $k$, of Algorithm~\ref{alg:RSD_RVO} is illustrated in Figure~\ref{fig:RSD_stage}.

\noindent
We next analyze Algorithm~\ref{alg:RSD_RVO} along two directions. First, contrary to 
Algorithm~\ref{alg:RSD_DVO}, the present algorithm may exit with 
a solution which is not $\epsilon$-probabilistic robust. This is due to the randomized nature
of the oracle, which may detect a ``false positive,'' by  misclassifying as good a solution $\theta^*_k$ for which 
$V(\theta^*_k)>\epsilon$. We  show that the probability of such a ``bad exit'' event can be made arbitrarily small.
Second, we fully characterize the probabilistic running time (iterations to exit) of the algorithm.
We start with the following key preliminary lemma, which is the backbone of the whole paper.

\vspace{.2cm}

\begin{lemma}
	\label{lem:main}
		Let Assumption~\ref{ass:unique} hold and, for any given  iteration $k$,
define the events
		\beas
		\mathrm{True} &=& \{ \mbox{$\epsilon'$-RVO returns {\tt true}}   \} \\
		\mathrm{GoodTrue} &=& \{ \mbox{$\epsilon'$-RVO returns {\tt true}} \cap V(\theta_k^*) \leq \epsilon   \} 		\\
			\mathrm{BadTrue} &=& \{ \mbox{$\epsilon'$-RVO returns {\tt true}} \cap V(\theta_k^*) > \epsilon   \} 	
		\eeas
Let
\beas
\lefteqn{ \fbb (N_o, n,N+1-n; i) } \\ && \doteq 
\binom{N_o}{i} \frac{B(i+n,N_o-i+N-n+1)}{B(n,N+1-n)} ,		\\
\lefteqn{H_{\epsilon,\epsilon'}(N,N_o)} \\ & &\doteq 1- \sum_{i=0}^{\lfloor \epsilon' N_o \rfloor} \fbb(N_o, n,N+1-n; i)\cdot \\ && \rule{1.5cm}{0cm} \cdot \FB  (n+i,N+N_o-n-i+1;\epsilon) ,\\
\lefteqn{H_{1,\epsilon'}(N,N_o)} \\ &&\doteq  1- \sum_{i=0}^{\lfloor \epsilon' N_o \rfloor} \fbb(N_o, n,N+1-n; i),\\
\lefteqn{
\bar\beta_{\epsilon,\epsilon'}(N,N_o) } \\ &&\doteq   
\FB(N+(1-\epsilon')N_o-n+1,n+\epsilon' N_o;1-\epsilon) . \label{eq:bbareps}
\eeas
At any iteration $k$ of Algorithm~\ref{alg:RSD_RVO}, it holds that
	\bea
	\lefteqn{
		\prob^{N+N_o} \{\mathrm{True}  \}   \geq  
 1-H_{1,\epsilon'}(N,N_o)},
	 \label{truebound} \\
\lefteqn{ \prob^{N+N_o} \{ \mathrm{GoodTrue} \}
 \geq  
1-H_{\epsilon,\epsilon'}(N,N_o) }
\label{eq:goodtruebound_exact} \\
&&\geq  (1-\bar\beta_{\epsilon,\epsilon'}(N,N_o)) (1-H_{1,\epsilon'}(N,N_o)).
\label{eq:goodtruebound} \\
\lefteqn{
\prob^{N+N_o} \{ \mathrm{BadTrue} \} }\label{eq:badtruebound}  \\ && \leq  
\FB((1-\epsilon')N_o, \epsilon' N_o+1;1-\epsilon) \beta_\epsilon (N).
\nonumber
	\eea
	Moreover, if
	problem (\ref{eq:rcp}) is f.s.\ w.p.\  one, then
	bounds  (\ref{truebound}) and (\ref{eq:goodtruebound_exact}) hold with equality, 
	and
	\bea
	\lefteqn{
	\prob^{N+N_o} \{ \mathrm{BadTrue} \} = H_{\epsilon,\epsilon'}(N,N_o) - H_{1,\epsilon'}(N,N_o)}\nonumber \rule{1.7cm}{0cm}\\
	&&\leq  \bar\beta_{\epsilon,\epsilon'}(N,N_o) (1-H_{1,\epsilon'}(N,N_o)). \rule{.6cm}{0cm}
	\label{eq:badtruebound_exact} 
	\eea
	\est
\end{lemma}

\noindent
See Section~\ref{app:lem:main} in the Appendix for a proof of Lemma~\ref{lem:main}.

\vspace{.2cm}
We can now state the main result concerning Algorithm~\ref{alg:RSD_RVO}.

\vspace{.2cm}

\begin{theorem}
	\label{thm:RSD_RVO}
		Let Assumption~\ref{ass:unique} hold. 
		Let  $\epsilon,\epsilon' \in [0,1]$, $\epsilon'\leq \epsilon$, and $N\geq n$ be given.
		Let 
		all the notation be set as in Lemma~\ref{lem:main},
			and let $\prob^{\times \times}$ denote the product probability
			$\prob^{N+N_o}\times \prob^{N+N_o}\times \cdots$.
			Define the event
			$\mathrm{BadExit}$ in which Algorithm~\ref{alg:RSD_RVO} exits returning a ``bad'' solution
			$\theta^*$: 
			\[
			\mathrm{BadExit} \doteq \{\mbox{Algorithm~\ref{alg:RSD_RVO} returns $\theta^*$: $V(\theta^*)>\epsilon$}\}.
			\]
 The following statements hold. 
\ben
\item 
\bea
\lefteqn{
\prob^{\times\times}\{\mathrm{BadExit}\} } \label{eq:badexitbound} \\
&&  \leq 
\frac{\FB((1-\epsilon')N_o, \epsilon' N_o+1;1-\epsilon)}{1-H_{1,\epsilon'}(N,N_o)} \beta_\epsilon (N).
\nonumber
\eea
If problem (\ref{eq:rcp}) is f.s.\ w.p.\  one, then it actually holds that
\beq
\prob^{\times\times}\{\mathrm{BadExit}\} \leq \bar\beta_{\epsilon,\epsilon'}(N,N_o)).
\label{eq:badexitbound_exact}
\eeq

\item The expected running time of Algorithm~\ref{alg:RSD_RVO} is $\leq (1-H_{1,\epsilon'}(N,N_o))\inv$, and equality holds if the scenario problem is f.s.\ w.p.\ 1.

\item The running time of Algorithm~\ref{alg:RSD_RVO} is 
$\leq k$ with probability
 $\geq 1- H_{1,\epsilon'}(N,N_o)^k$, and equality holds if the scenario problem is f.s.\ w.p.\ 1.
\een
\est
\end{theorem}

\noindent
See Section~\ref{app:thm:main} in the Appendix for a proof of Theorem~\ref{thm:RSD_RVO}.

\vspace{.2cm}
\subsubsection{Asymptotic bounds}
\label{sec:approxbound}
A key quantity related to the expected running time of Algorithm~\ref{alg:RSD_RVO}
is $H_{1,\epsilon'}(N,N_o)$, which is the upper tail of a beta-Binomial distribution.
This quantity is related to the hypergeometric function $_3F_2$, and to ratios of Gamma functions, which may be delicate to evaluate numerically for large values of the arguments.
It is therefore useful to have a more ``manageable,'' albeit approximate, expression for $H_{1,\epsilon'}(N,N_o)$. The following corollary gives an asymptotic expression for
$H_{1,\epsilon'}(N,N_o)$, see Section~\ref{app:cor:approxbound} in the Appendix for a proof.
\begin{corollary}
\label{cor:approxbound}
For $N_o\to \infty$ it holds that
\beq
H_{1,\epsilon'}(N,N_o) \to  \beta_{\epsilon'}(N).
\label{eq:H1_bound}
\eeq
%
\est
\end{corollary}
An interesting consequence of Corollary~\ref{cor:approxbound} is that, for large $N_o$,
and $\epsilon' \leq \epsilon$, we have $H_{1,\epsilon'}(N,N_o) \simeq  \beta_{\epsilon'}(N) \geq \beta_{\epsilon}(N)$, from which we conclude that
\beq
\hat K \doteq \frac{1}{1-H_{1,\epsilon'}(N,N_o)} \simeq  \frac{1}{1-\beta_{\epsilon'}(N)} \geq  \frac{1}{1-\beta_{\epsilon}(N)}  .
\label{eq:exprunning_bound}
\eeq
This last equation gives us
an approximate, asymptotic,  expression 
for 
 the upper bound $\hat K$  on the expected running time of Algorithm~\ref{alg:RSD_RVO}, and
also tells us that, for  $\epsilon' \leq \epsilon$, this bound cannot be better (smaller) 
 than the corresponding bound of  the
``ideal'' Algorithm~\ref{alg:RSD_DVO}.

\subsection{Practical dimensioning of the scenario and oracle blocks}
\label{sec:dimensioning}
In a typical probabilistic design problem we are given the dimension $n$ of the decision variable and
the level $\epsilon\in(0,1)$ of probabilistic robustness we require from our design.
If we intend to use a randomized approach, we also set a confidence level $1-\beta\in(0,1)$,
which is the a-priori level of probability with which our randomized approach will be successfull in returning an $\epsilon$-probabilistic robust design.
In a plain (i.e., non repetitive) scenario design setting, this requires dimensioning the
number $N$ of scenarios so to guarantee that $\beta_\epsilon(N) \leq \beta$; this can be done, for instance, by using the bound in (\ref{eq:Nbound}), or via a simple numerical search over $N$.
However, if the required $N$ turns out to be too large in practice (e.g., the ensuing scenario optimization problem becomes impractical to deal with numerically), we can switch to a repetitive scenario design approach. 
In such a case, we suggest the following route for designing the scenario and oracle blocks.
Let us first select a level $\epsilon' \leq \epsilon$ to be used in the oracle.
Qualitatively, decreasing $\epsilon'$ increases the expected running time and decreases the required $N_o$,
and the converse happens for increasing $\epsilon'$. We here suggest to set $\epsilon'$ in the range
$ [0.5, 0.9]\epsilon$.

\subsubsection{Dimensioning the scenario block}
We 
 dimension the scenario optimization block by choosing $N$
 so to achieve a good tradeoff
 between the complexity of the scenario program (which grows with $N$)
 and the expected number of iterations required by the RSD approach (which decreases with $N$). 
 This choice can be made, for instance, by plotting the approximate expression
 (which becomes exact as $N_o\to \infty$)
  in (\ref{eq:exprunning_bound})
for the upper bound on the expected running time of
 Algorithm~\ref{alg:RSD_RVO}, $(1-\beta_{\epsilon'}(N))\inv$, versus $N$, and selecting a value of $N$ for which
 this running time is acceptable.

 \subsubsection{Dimensioning the oracle block}
 Once $N$ has been selected according to the approach described above, we
consider point 1 and point 2 in Theorem~\ref{thm:RSD_RVO} and 
we dimension the $\epsilon'$-RVO block by searching numerically for an $N_o$ 
%
such that
 the  right-hand side of (\ref{eq:badexitbound}) (or of (\ref{eq:badexitbound_exact}), if the problem is f.s.) is $\leq \beta$.

\begin{remark}\rm
\label{rem:bounds}
We observe that, in general, the bound in (\ref{eq:badexitbound}) should be used
for the design of the $\epsilon'$-RVO block. However, 
 the expression in (\ref{eq:badexitbound_exact}) is easier to deal with than the one in (\ref{eq:badexitbound}).
 It is hence
 advisable to use the former in a preliminary dimensioning phase; the so-obtained values
 can then be verified ex-post against the actual bound in (\ref{eq:badexitbound}).
 Another advantage of (\ref{eq:badexitbound_exact})
 is that, using a bounding technique analogous to the one described in Section~5 of \cite{Calafiore:SIOPT10},
we can ``invert'' the condition $\bar\beta_{\epsilon,\epsilon'}(N,N_o)\leq \beta$,
%
%
finding (after some manipulation) that this condition is satisfied if
\beq
N_o \delta + N(\delta/2 + \epsilon') \geq \frac{\epsilon}{\delta}\ln\beta\inv + n - 1,\quad
\delta \doteq \epsilon - \epsilon' >0.
\label{eq:NNobound}
\eeq
With a choice of the pair $(N,N_o)$ such that
(\ref{eq:NNobound}) is satisfied, 
 we guarantee a priori that our randomized Algorithm~\ref{alg:RSD_RVO} may fail in returning an $\epsilon$-probabilistic robust design
w.p.\  at most $\beta$, as desired (rigorously, this only holds under the assumption that the scenario problem is
f.s.\ w.p.\ one).
The nice feature highlighted by  (\ref{eq:NNobound}) is that now the ``workload'' necessary to achieve the desired failure level $\beta$
is subdivided between $N$ (samples in the scenario problem) and 
$N_o$ (samples in the oracle): a lower complexity scenario problem can be employed, as long as
it is paired with a randomized oracle having a suitable $N_o$. Notice, however, that,
in making the choice of the $(N,N_o)$ pair, the expected running time of 
Algorithm~\ref{alg:RSD_RVO} should also taken into consideration, and that this places
a lower limit on how small $N$ can be, 
see also the discussion in Section~\ref{sec:approxbound}.
\end{remark}

\begin{remark}\rm
\label{rem:parallel}
We further observe that, in typical cases, dealing with large $N_o$ is a milder problem
than dealing with large $N$. This is due to the fact that merely {\em checking} 
satisfaction of inequality $f(\theta^*_k, q^{(i)})>0$, for $i=1,\ldots,N_o$, is generally easier than 
solving a related optimization problem with as many constraints. 
Also, we remark that the $\epsilon'$-RVO algorithm is inherently parallel, so an $M$-fold speedup can  potentially be gained if $M$ processors are available in parallel for the randomized feasibility test.
 Actually, the whole approach can be formulated in a fully parallel -- instead of sequential -- form, where $W$
 workers solve in parallel $W$ instances of scenario problems, and each 
worker has its own $M$ parallel sub-workers to be used in the randomized oracle.
Such a parallel version of the RSD method can be easily analyzed using the probabilistic tools developed in this paper.

\end{remark}

\section{Numerical examples}
\label{sec:examples}
We exemplify the steps of the RSD approach, from algorithm dimensioning to numerical results,
using two  examples of robust control design. The first example deals with
robust finite-horizon input design for an uncertain linear system, while the second example
deals with robust performance design for a positive linear system.

\subsection{Robust finite-horizon input design}
\label{sec:es:mpc}
We consider a system of the form
\[
x(t+1) = A(q)x(t) + B u(t),\quad t=0,1,\ldots;\; x(0) = 0,
\] 
where 
$u(t)$ is a scalar input signal, and
$A(q)\in\Real{n_a,n_a}$ is an interval uncertain matrix of the form
\[
A(q) = A_0 + \sum_{i,j=1}^{n_a} q_{ij} e_i e_j\tran,\quad
|q_{ij}|\leq \rho,\; \rho>0,
\]
where $e_i$ is a vector of all zeros, except for a one in the $i$-th entry.
%
Given a final time $T\geq 1$ and a target state $\bar x$, the problem is to determine
an input sequence $\{u(0),\ldots,u(T-1)\}$ such that {\em (i)} the state $x(T)$ is robustly contained in a small ball around the target state $\bar x$, and {\em (ii)} the input energy $\sum_k u(k)^2$ is not too large.
We write
$
x(T) = x(T; q) = \calR (q) u$, where
$ \calR (q) $ is the $T$-reachability matrix of the system (for a given $q$), 
and $u\doteq (u(0),\ldots,u(T-1))$. Then,
we formally express our design goals 
in the form of minimization of a level $\gamma$ such that
\[
\| x(T; q) -\bar x \|_2^2 + \lam \sum_{t=0}^{T-1} u(t)^2 \leq \gamma ,
\]
where $\lam\geq 0$ is a tradeoff parameter. Letting $\theta = (u,\gamma)$, the problem is formally stated in our framework by setting
\[
f(\theta,q) \leq 0,\quad \mbox{where } f(\theta,q)\doteq \|\calR (q) u - \bar x\|_2^2 + \lam \|u\|_2^2-\gamma.
\]
Assuming that the uncertain parameter $q$ is random and uniformly
distributed in the hypercube $\mathbb{Q} = [-\rho,\rho]^{n_a\times n_a}$, 
our scenario design problem takes the following form:
\beas
\min_{\theta= (u,\gamma)} & \gamma \\
\mbox{s.t.:} & f(\theta,q^{(i)}) \leq 0,\quad i=1,\ldots,N.
\eeas

\paragraph{Dimensioning the RSD algorithm}
We set $T = 10$, thus
the size of the decision variable $\theta = (u, \gamma)$
of the scenario problem is $n = 11$.
%
We set the desired level of probabilistic robustness
 to $1-\epsilon = 0.995$, i.e., $\epsilon = 0.005$, and require a level of failure of the randomized method below
$\beta = 10^{-12}$, that is, we require the randomized method to return a good solution with ``practical certainty.''
Using a plain (one-shot) scenario approach, imposing  $\beta_\epsilon(N)\leq \beta$ would require
 $N\geq 10440$ scenarios.
Let us now see how we can reduce this $N$ figure by resorting to a repetitive scenario design approach.

\begin{figure}[htb]
\begin{center}
\includegraphics[width=.5\textwidth]{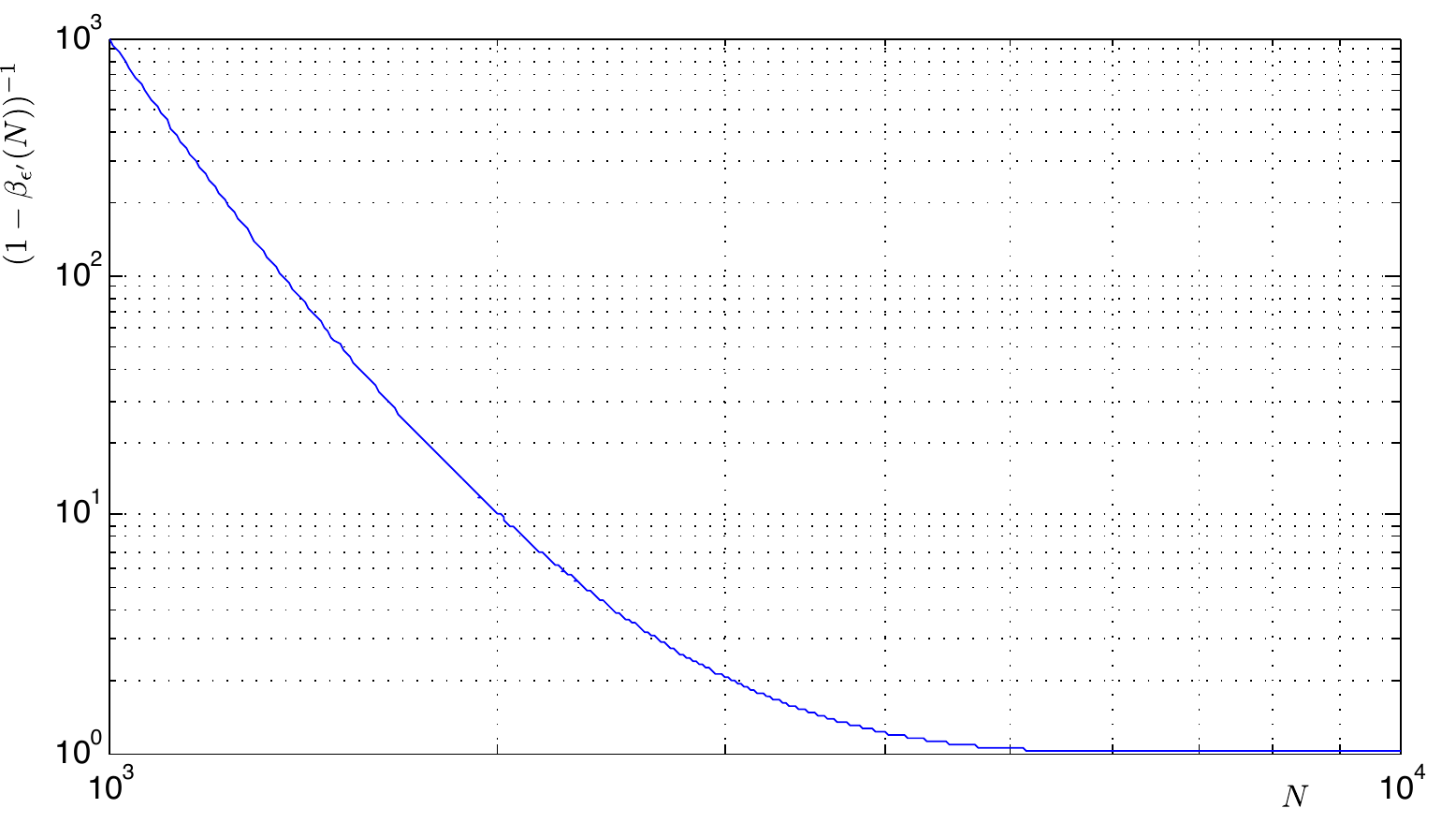}
\caption{Example in Section~\ref{sec:es:mpc}:  Log-log plot of $(1-\beta_{\epsilon'}(N))\inv$ vs.\ $N$. \label{fig:tradeoffcurve} }%
\end{center}%
\end{figure}

Let us fix $\epsilon' = 0.7\epsilon =  0.0035$, thus $\delta = \epsilon-\epsilon' = 0.0015$.
A plot of the (asymptotic) bound on expected number of iterations, $(1-\beta_{\epsilon'}(N))\inv$, as a function of $N$ is shown in Figure~\ref{fig:tradeoffcurve}. We see from this plot, for instance, that the choice
$N = 2000$ corresponds to a value of about $10$ for the upper bound on the   expected number of iterations in Algorithm~\ref{alg:RSD_RVO}.
Let us choose this value of $N$ for the scenario block.

For $\beta = 10^{-12}$, the simplified condition in (\ref{eq:NNobound}) tells us that
$N_o \geq 62403$. Let us choose $N_o = 63000$ samples to be used in the oracle.
With the above choices we have
 $H_{1,\epsilon'}(N,N_o) = 0.8963$,
thus the algorithm's upper bound on average running time is $\hat K = (1-H_{1,\epsilon'}(N,N_o))\inv = 9.64$. Notice that this upper bound is tight for f.s.\ problems, but
it is conservative for problems that are not necessarily f.s. Thus, in general, we may expect a performance which is in practice better than the one predicted by
the theoretical worst-case bound.
%
 
\paragraph{Numerical test}
\begin{figure*}[!!!t]
	\normalsize
\begin{equation*}
A_0 = \left[\begin{array}{cccccc} -0.7214 & -0.0578 & 0.2757 & 0.7255 & 0.2171 & 0.3901\\ 0.5704 & 0.1762 & 0.3684 & -0.0971 & 0.6822 & -0.5604\\ -1.3983 & -0.1795 & 0.1511 & 1.0531 & -0.1601 & 0.9031\\ -0.6308 & -0.0058 & 0.4422 & 0.8169 & 0.512 & 0.2105\\ 0.7539 & 0.1423 & 0.2039 & -0.3757 & 0.5088 & -0.6081\\ -1.3571 & -0.1769 & 0.1076 & 1.0032 & -0.1781 & 0.9151 \end{array}\right],
\quad B=\left[\ba{c} 0 \\1\\ 0\\ 1\\ 0\\ 1\ea\right]
\end{equation*}
\hrulefill
\vspace*{4pt}
\end{figure*}

\begin{figure*}[h!t]
\begin{center}
\includegraphics[width=1\textwidth]{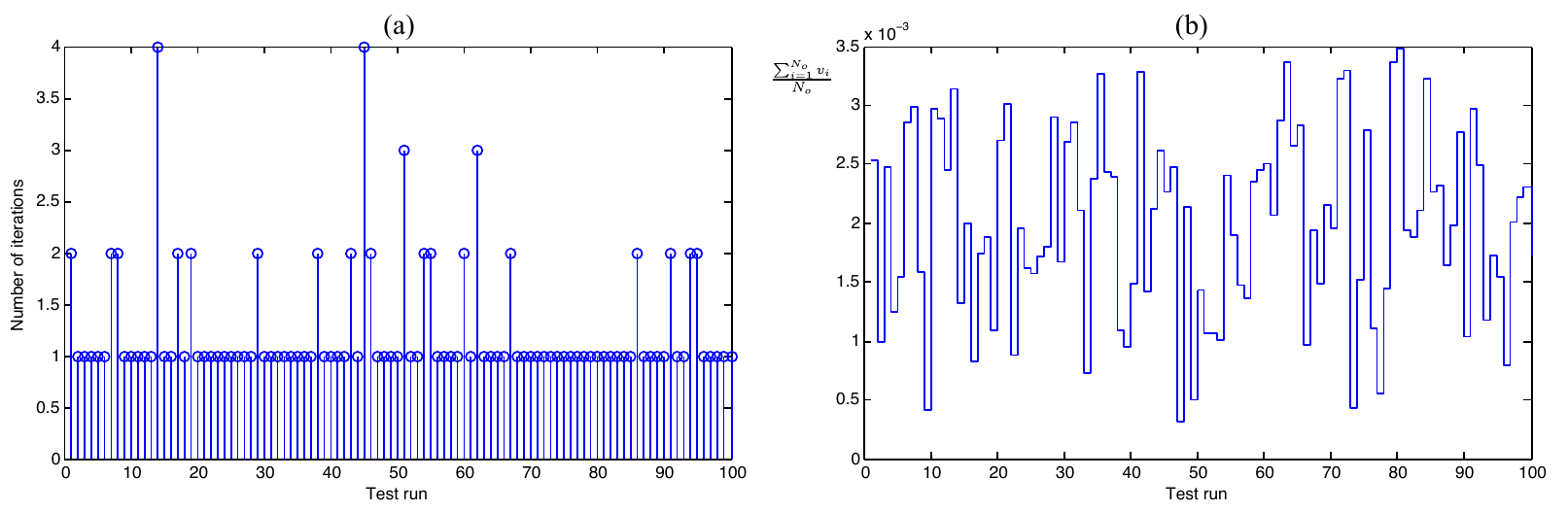}
\caption{Example in Section~\ref{sec:es:mpc}: (a) Repetitions of
Algorithm~\ref{alg:RSD_RVO} in the  $100$ test runs.
(b) Levels of empirical violation probability evaluated by the oracle upon exit,
 in the  $100$ test runs.
\label{fig:iters} }%
\end{center}%
\end{figure*}

\begin{figure*}[h!t]
\begin{center}
\includegraphics[width=1\textwidth]{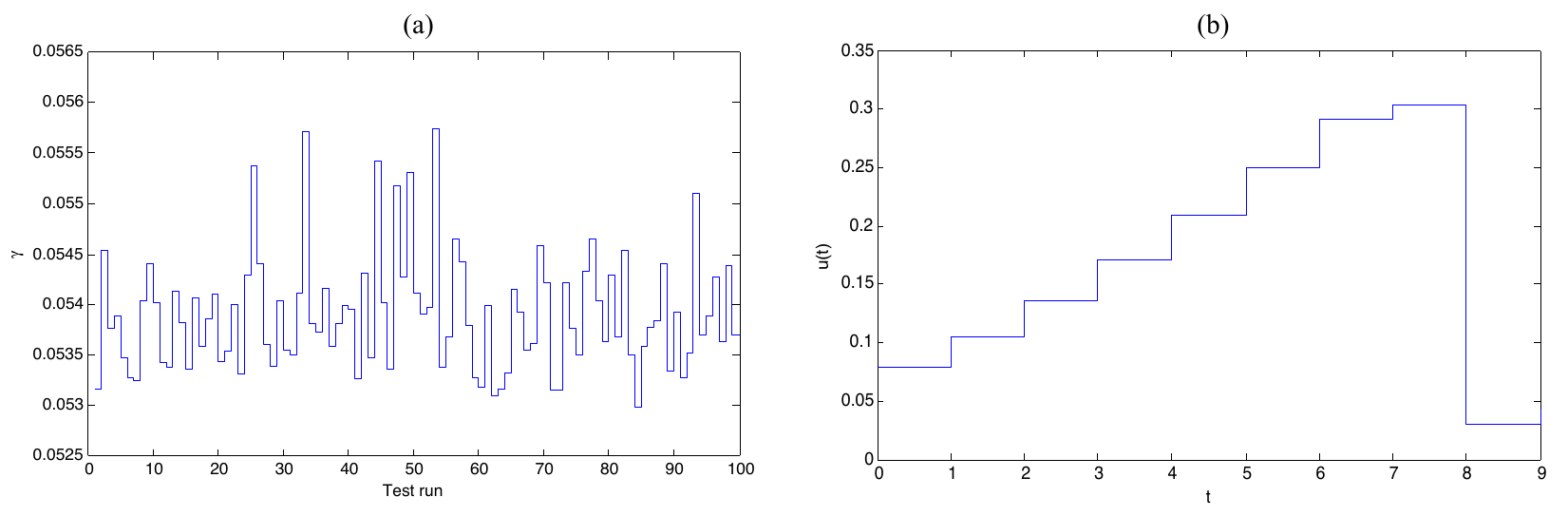}
\caption{Example in Section~\ref{sec:es:mpc}: (a) Optimal $\gamma$ level returned 
by
Algorithm~\ref{alg:RSD_RVO} in the  $100$ test runs.
(b) 
Average over the 100 test runs of the optimal
input $u(t)$ returned 
by
Algorithm~\ref{alg:RSD_RVO}.
\label{fig:gamma} }%
\end{center}%
\end{figure*}

We considered
the nominal matrix $A_0$ of dimension $n_a=6$ and $B$ matrix shown on top of this page,
with target state $\bar x =[1, -1/2, 2, 1, -1, 2]\tran$, 
$\rho = 0.001$, and $\lam = 0.005$.
We run Algorithm~\ref{alg:RSD_RVO} for $100$ times, and on each test run we recorded the number of iterations and the solution returned upon exit.
Figure~\ref{fig:iters}(a) shows the number of repetitions in the test runs: we see that the algorithm exited most of the times in a single repetition, with a maximum of 4 repetitions, which is below the figure predicted by the upper bound $\hat K =  9.64$: practical performance was thus better than predicted, which suggests that the problem at hand is not fully supported w.p.\ 1.
Figure~\ref{fig:iters}(b)  shows the level of empirical violation probability evaluated by the oracle upon exit.
Finally, Figure~\ref{fig:gamma}(a) shows the optimal $\gamma$ level
returned by the algorithm in the test runs, and Figure~\ref{fig:gamma}(b)
shown the optimal input signal returned by the algorithm, averaged over the 100 test runs.

\paragraph{Computational improvements}
In this example, the RSD approach permitted a substantial reduction
of the number of design samples (from the 10440 samples required by the
plain scenario method, to just 2000 samples), at the price of a very moderate number of repetitions (the average number of repetitions in the 100 test runs was 1.27). 

The numerical experiments were carried out on an Intel Xeon X5650 machine using CVX under Matlab; \cite{cvx}.
On average over the 100 test experiments, the RSD method (with $N=2000$, $N_o=63000$) required $224$ s to return a solution. For comparison purposes, we also run a plain, one-shot, scenario optimization with the $N=10440$ scenarios that are required to attain the desired $\beta=10^{-12}$ level: the time required for obtaining such a solution was
$2790$ s. Using the RSD approach instead of a plain one-shot scenario design thus yielded 
a reduction in computing time of about one order of magnitude.
The reason for this improvement is due to the fact that the scenario optimization problem  in the RSD approach
(which uses $N=2000$ scenarios)  took about $173$ s to be solved on a typical run, and the subsequent randomized oracle test (with $N_o=63000$) is computationally cheap, taking only about $3.16$ s. 

\subsection{An uncertain linear transportation network}
\label{sec:es:transport}
\begin{figure}[b!!t]
\begin{center}
\includegraphics[width=.3\textwidth]{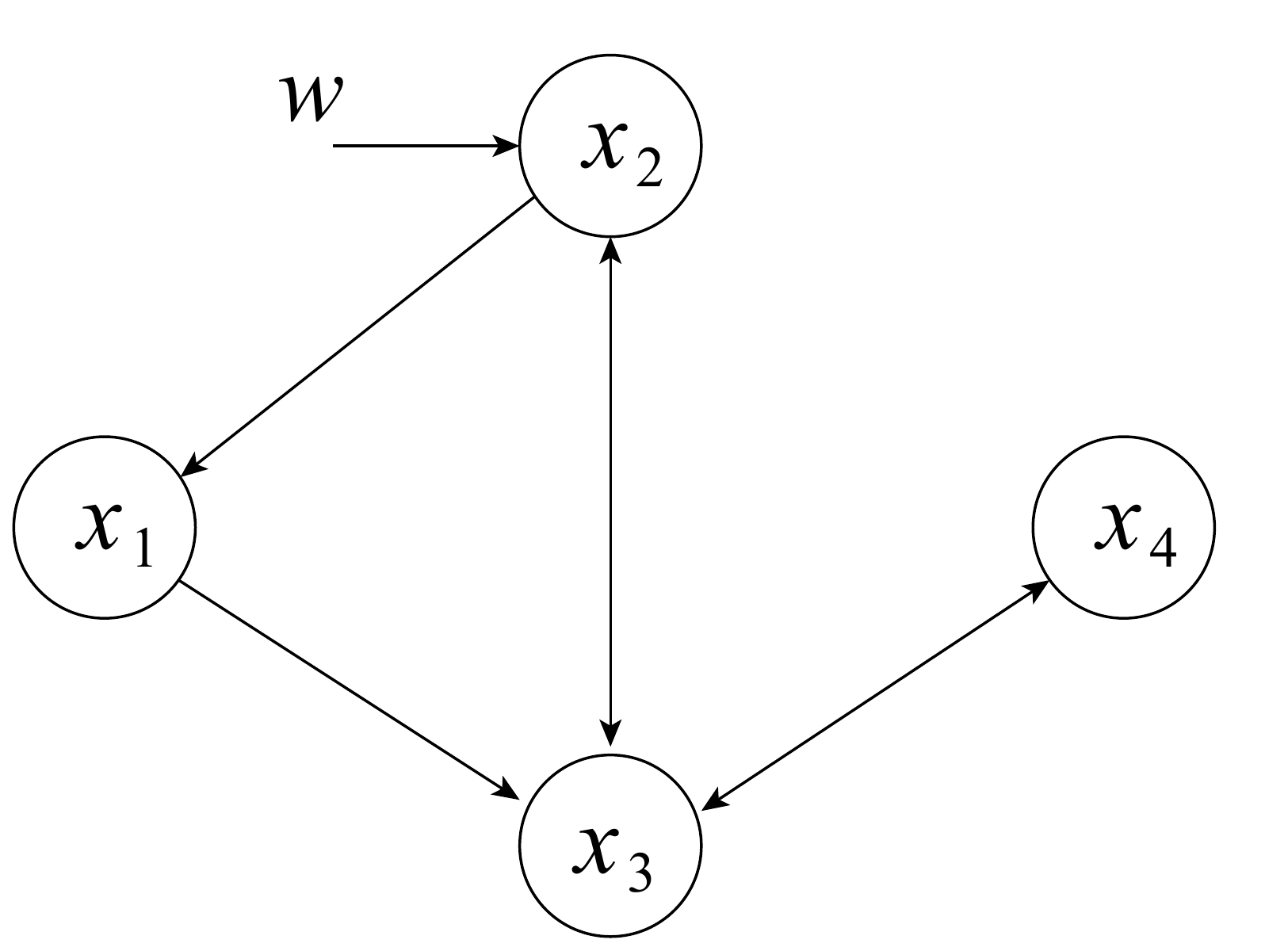}
\caption{Example in Section~\ref{sec:es:transport}:   A network model.
\label{fig:network} }%
\end{center}%
\end{figure}

As a second example, we consider a variation on a transportation network model introduced in Section~3 of \cite{Rantzer:14}; see Figure~\ref{fig:network}.

The model is described by the state equations where
the states $x_i$, $i=1,\ldots,4$, represent the contents of four buffers,  the parameters $\ell_{ij}\geq 0$ represent
the rate of transfer from buffer $j$ to buffer $i$,  $w(t)\geq 0$ is an input flow on the second buffer, and we take as output $y$
the total content of the buffers; see eq.\ (\ref{eq:net-x})-(\ref{eq:net-y}).

 \begin{figure*}[!!!t]
\bea
\dot{x} &=& \left[ \ba{cccc} 
-1-\ell_{31} & \ell_{12} & 0 & 0 \\ 0 & -\ell_{12}-\ell_{32} & \ell_{23} & 0 \\
 \ell_{31} &  \ell_{32} &  -\ell_{23}- \ell_{43} &  \ell_{34} \\
 0 & 0 &  \ell_{43} & -4-\ell_{34} 
\ea\right] x + 
\left[ \ba{c} 
0 \\ 1 \\ 0 \\ 0
\ea\right] w  \label{eq:net-x}\\
y &=& \left[ \ba{cccc} 
1 & 1 & 1 & 1 \ea\right] x. \label{eq:net-y}
\eea
\end{figure*}

We consider the situation in which
 $\ell_{31} = 2+q_1$, $\ell_{34} = 1+q_2$, $\ell_{43}=2+q_3$, where
 $\ell=[\ell_{12}\;\ell_{23}\;\ell_{32}]\tran \in[0,1]^3$ is a vector of parameters to be designed, and
 $q =[q_1\; q_2\; q_3]\tran$ is an uncertainty term, which is assumed to be  a truncated 
 Normal random vector with zero mean, covariance matrix $\Sigma = 0.2^2 I$, and
 $\|q\|_\infty \leq 1$.
This system has the form $\dot{x} = A(\ell,q) x + B w$, $y=C x$, where $B\geq 0$,  $C\geq 0$ (element-wise), and
the $A(\ell,q)$ matrix is Metzler (i.e., the off-diagonal entries of $A$ are nonnegative). Theorem~4 in \cite{Rantzer:14} states that, for given $\ell, q$,
this system is stable and the peak-to-peak gain from $w$ to $y$ is smaller that some given $\gamma$ if and only if
there exist $\xi \geq 0$ such that
\[
\left[ \ba{cc} 
A(\ell,q) & B \\ C & 0 
\ea\right]
\left[ \ba{c} 
\xi \\ \ones  
\ea\right] < 
\left[ \ba{c} 
0\\ \gamma\ones 
\ea\right],
\]
where $\ones$ is a vector of ones.
 By taking $N$ i.d.d.\ samples $q^{(i)}$ of $q$, a robust  scenario design problem is one in which one seeks to minimize the  
 peak-to-peak gain $\gamma$ subject to
 the above constraint on the scenarios; see eq.\ (\ref{eq:es2:scenario0}).
  \begin{figure*}[!!!t]
	\normalsize
 \bea
 \min_{\ell_{12},\ell_{23},\ell_{32} \in[0,1]; \gamma\; \xi \geq 0} & \gamma \label{eq:es2:scenario0} \\
 \mbox{s.t.:} & 
 \left[ \ba{cccc} 
-3-q_1^{(i)} & \ell_{12} & 0 & 0 \\ 
0 & -\ell_{12}-\ell_{32} & \ell_{23} & 0 \\
 2+q_1 &  \ell_{32} &  -\ell_{23}- 2-q_3^{(i)} &  1+q_2^{(i)} \\
 0 & 0 &  2+q_3^{(i)} & -5- q_2^{(i)}
\ea\right] \xi + B\ones  < 0, & i=1,\ldots,N \nonumber \\
& C\xi < \gamma\ones. \nonumber
 \eea
 \hrulefill
\vspace*{4pt}
\label{fig:eqscen}
\end{figure*}
This problem is a ``robustified'' version of the one discussed in Section~V of \cite{Rantzer:14}.
The problem as stated is not convex, due to the product terms between entries in $\xi$ and $\ell$. However, by introducing new variables
$\mu_{12} = \ell_{12}\xi_2$, $\mu_{32} = \ell_{32}\xi_2$, $\mu_{23} = \ell_{23}\xi_3$, we rewrite the problem as an LP in the variables $\xi$, $\mu = [\mu_{12}\; \mu_{32}\; \mu_{23}]\tran$, and $\gamma$; see
eq.\ (\ref{eq:es2:scenario}).
 \begin{figure*}[!!!t]
	\normalsize
 \bea
 \min_{\mu\geq 0; \xi \geq 0; \gamma} & \gamma \label{eq:es2:scenario}\\
 \mbox{s.t.:} & 
 \left[ \ba{cccc} 
-3-q_1^{(i)} & 0 & 0 & 0 \\ 
0 & 0 & 0 & 0 \\
 2+q_1^{(i)}  &  0 &  - 2-q_3^{(i)} &  1+q_2^{(i)} \\
 0 & 0 &  2+q_3^{(i)} & -5- q_2^{(i)}
\ea\right] \xi 
+
 \left[ \ba{ccc} 
1 & 0 & 0  \\ 
-1 & -1 & 1  \\
 0 & 1 & -1 \\
 0 & 0 & 0\ea\right] 
 \mu
+ B\ones  < 0, & i=1,\ldots,N \nonumber \\
& C\xi < \gamma\ones,\quad
 \mu_{12}\leq \xi_{2},\; 
\mu_{32}\leq \xi_{2}, \;
\mu_{23}\leq \xi_{3}.\nonumber
 \eea
 \hrulefill
\vspace*{4pt}
\label{fig:eqscen2}
\end{figure*}

\paragraph{Dimensioning the RSD algorithm}
 The size of the decision variable $\theta = (\xi, \mu, \gamma)$
of the scenario problem is $n = 8$.
As in the previous example, we set the desired level of probabilistic robustness
 to $1-\epsilon = 0.995$, i.e., $\epsilon = 0.005$, and require a level of failure of the randomized method below
$\beta = 10^{-12}$.
Using a plain (one-shot) scenario approach, imposing  $\beta_\epsilon(N)\leq \beta$ would require
 $N\geq  9197$ scenarios.
We next reduce this $N$ figure by resorting to a repetitive scenario design approach.

Let us fix $\epsilon' = 0.7\epsilon =  0.0035$, thus $\delta = \epsilon-\epsilon' = 0.0015$.
Plotting  the asymptotic bound on expected number of iterations, $(1-\beta_{\epsilon'}(N))\inv$ as a function of $N$ 
(as we did in Figure~\ref{fig:tradeoffcurve} for the previous example), we see that
 the choice
$N = 1340$ corresponds to a value of about $10$ for the upper bound on the   expected number of iterations in Algorithm~\ref{alg:RSD_RVO}.
Let us choose this value of $N$ for the scenario block.

For $\beta = 10^{-12}$, the simplified condition in (\ref{eq:NNobound}) tells us that
$N_o \geq 62273$ samples can be used in the randomized feasibility oracle.
With the above choices we have
 $H_{1,\epsilon'}(N,N_o) = 0.8931$,
thus the algorithm's upper bound on average running time is $\hat K = (1-H_{1,\epsilon'}(N,N_o))\inv = 9.36$ (notice again that, in general, we may expect a performance which is in practice better than the one predicted by
this theoretical worst-case bound, since the the actual problem may not be fully supported).

\begin{figure*}[h!tb]
\begin{center}
\includegraphics[width=1\textwidth]{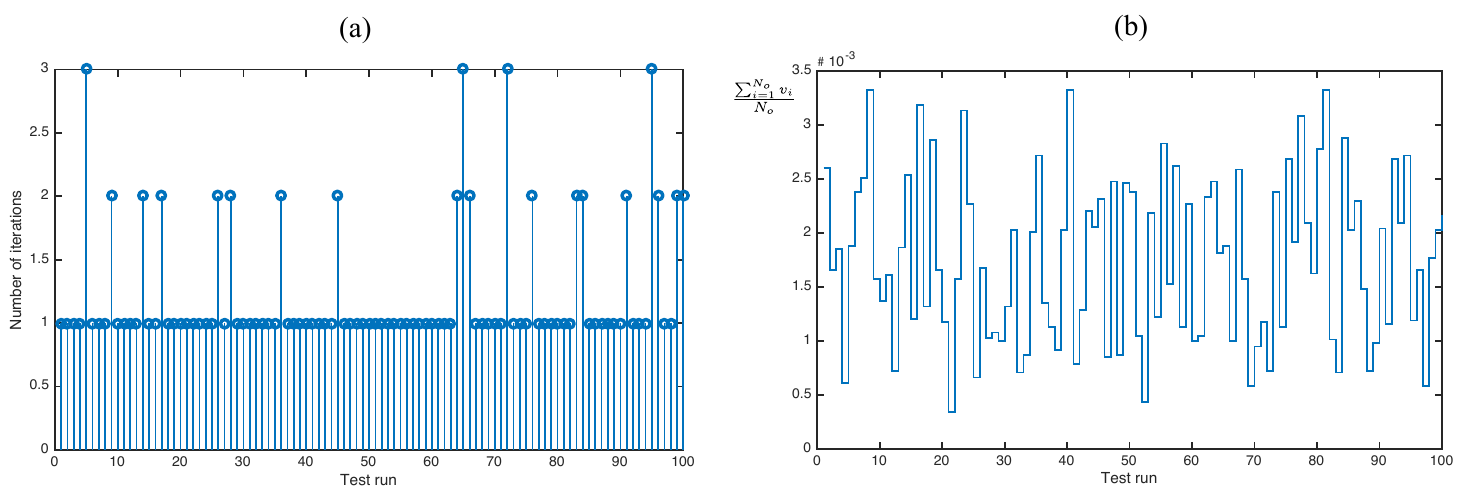}
\caption{Example in Section~\ref{sec:es:transport}: (a) Repetitions of
Algorithm~\ref{alg:RSD_RVO} in the  $100$ test runs.
(b) Levels of empirical violation probability evaluated by the oracle upon exit,
 in the  $100$ test runs.
\label{fig:iters2} }%
\end{center}%
\end{figure*}

\begin{figure}[h!t]
\begin{center}
\includegraphics[width=.5\textwidth]{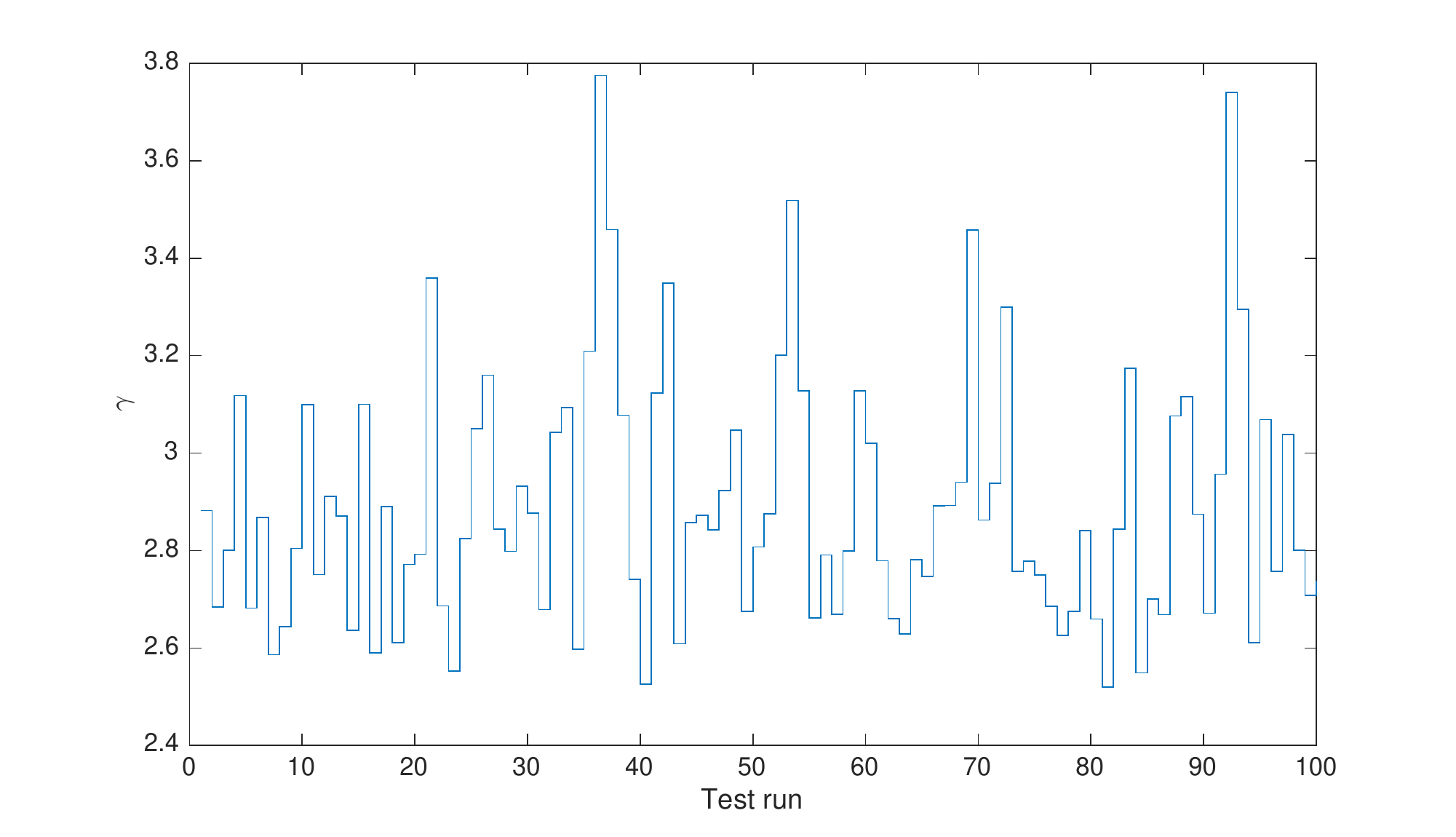}
\caption{Example in Section~\ref{sec:es:transport}:   Optimal $\gamma$ level returned 
by Algorithm~\ref{alg:RSD_RVO} in the  $100$ test runs.
\label{fig:gamma2} }%
\end{center}%
\vspace{-.5cm}
\end{figure}

\paragraph{Numerical test and computational performance}
We first solved the problem via a plain scenario approach, using $N =  9197$ scenarios.
The computational time was of about $50$ s, resulting in the
following optimal solution:
\[
\xi = \left[\ba{c}   0.2314\\
    0.5000\\
    1.7206\\
    0.9763 \ea\right], \quad 
 \mu = \left[\ba{c}   0.5000\\
    0.5000\\
    0.0000 \ea\right], \quad    
   \gamma =  3.4283.
\]
Next, we run the RSD method (Algorithm~\ref{alg:RSD_RVO}, with $N = 1340$, $N_o = 62273$) for $100$ times, and on each test run we recorded the number of iterations and the solution returned upon exit.
Figure~\ref{fig:iters2}(a) shows the number of repetitions in the test runs: we see that the algorithm exited most of the times in a single repetition, with a maximum of 3 repetitions; average 1.24 repetitions.
Figure~\ref{fig:iters2}(b)  shows the level of empirical violation probability evaluated by the oracle upon exit.
Finally, Figure~\ref{fig:gamma2} shows the optimal $\gamma$ level
returned by the algorithm in the test runs.

The average (over the 100 test trials) running time of the RSD method was about $6.4$ s.
Since the plain scenario approach required about
50 s, it  was about $680 \%$ slower than
the newly proposed RSD approach, in this test example.
Each repetition of the RSD method required about 
$4.6$ s for solving the scenario problem (with $N = 1340$), and $0.6$ s for the randomized oracle check
(with $N_o = 62273$); once again, we observe that the oracle time was much lower than the scenario optimization time.

\section{Conclusions}
Repetitive scenario design generalizes the scenario approach to robust design
by setting up an iterative
procedure whereby scenario design {\em trials} are followed by  a randomized check on the feasibility level of the solution. The expected number  of repetitions (or trials) in this procedure is dictated by the key quantity $H_{1,\epsilon'}(N,N_o)$, which is well approximated, for large $N_o$, by $\beta_{\epsilon'}(N)$. For $H_{1,\epsilon'}(N,N_o)\to 0$ we recover the extreme situation of the standard, one-shot, scenario design, in which a valid solution is found in a single repetition, at the cost of possibly large $N$. For smaller $N$ values, we can trade off complexity in the solution of the scenario problem for additional iterations in the RSD algorithm.
The extent to which $N$ can be reduced is however limited by the upper bound  $\hat K$ we impose on the expected running time, since eq.\ (\ref{eq:exprunning_bound}) tells us that
$H_{1,\epsilon'}(N,N_o) \simeq \beta_{\epsilon'}(N) \leq 1- \hat K\inv$.
Numerical examples showed that the proposed RSD approach may lead to improvements in computational time of about one order of magnitude, compared to a plain scenario approach.

\appendix
\small
\section{Appendix}

\subsection{Proof of Theorem~\ref{thm:RSD_DVO}}
\label{app:thm:RSD_DVO}
The first point of the theorem is obvious, 
sice the algorithm terminates if and only if {\tt true} is returned by the deterministic oracle,
which happens if and only if the condition $V(\theta^*_k) \leq \epsilon$ is satisfied.

For point two,
let $z_k= z_k(\omega^{(k)})$, $k=1,\ldots$, be i.i.d.\ Bernoulli variables
representing the outcome of the $\epsilon$-DVO step at each iteration, i.e.,
 $z_k=1$ if $V(\theta^*_k)\leq \epsilon$ (oracle returns {\tt true}), and $z_k=0$ otherwise
(oracle returns {\tt false}). 
From eq.\ (\ref{eq:betabound}) we observe that the probability of $z_k=1$ is $F_V(\epsilon)  \geq 1-\beta_\epsilon(N)$.
Since the algorithm terminates as soon as  a {\tt true}
is returned by the oracle, the {\em running time} of the algorithm is defined as the random variable
\[
K \doteq \{\mbox{\small iteration $k$ at which {\tt true} is returned for the first time}\}.
\]
Clearly, $K$ has a geometric distribution
\[
\prob^{\times} \{K = k \} = (1-F_V(\epsilon))^{k-1}F_V(\epsilon),
\]
where $\prob^{\times}$ denotes the product probability measure over $\omega^{(1)},\omega^{(2)},\ldots$
The mean of this geometric distribution is $1/F_V(\epsilon)$, whence
\[
\Ev \{K\} = \frac{1}{F_V(\epsilon)} \leq \frac{1}{1-\beta_\epsilon(N)},
\]
which proves the second point (note that equality holds if the scenario problem is f.s.\ w.p.\ one).
The cumulative of the above geometric distribution is
\[
\prob^{\times} \{K \leq k \} =  1-   (1-F_V(\epsilon))^{k}.
\]
This function is increasing in $F_V(\epsilon)$, thus $F_V(\epsilon)  \geq 1-\beta_\epsilon(N)$
implies
\[
\prob^{\times} \{K \leq k \} \geq  1-   \beta_\epsilon(N)^{k},
\]
which proves the third point. 
\qed

\subsection{Proof of Lemma~\ref{lem:main}}
\label{app:lem:main}
At any given iteration $k$ of Algorithm~\ref{alg:RSD_RVO}, let us consider the sequence
of binary random variables appearing inside the $\epsilon'$-RVO:
\[
v_i = \left\{\ba{ll}
1 & \mbox{if } f(\theta^*_k,q^{(i)}) > 0 \\
0 & \mbox{otherwise},
 \ea \right. \quad i=1,\ldots,N_o
\]
By definition, we have that $\prob \{q: f(\theta^*_k,q) > 0\} = V(\theta^*_k)$,
and $V(\theta^*_k)$ is a random variable with cumulative distribution function given by $F_V$.
Therefore, for {\em given}  $V(\theta^*_k) = p$, the $v_i$s form an i.i.d.\ Bernoulli sequence
with success probability $p$. However, $p$ is itself a random variable having cumulative distribution
$F_V$. Therefore, the $v_i$s form a so-called {\em conditionally i.i.d.}\ Bernoulli sequence
\cite{Aldous:83}, having $F_V$ as the directing de Finetti measure. In simpler terms, the $v_i$s are described by a {\em compound} distribution: first a success probability $p$ is extracted at random according to its
directing distribution $F_V$, and then the $v_i$s are generated according to an i.i.d.\ Bernoulli distribution with success probability $p$.
Let 
$
S \doteq \sum_{i=1}^{N_o} v_i
$.
Conditional on $V(\theta^*_k) = p$, the random variable $S$ has Binomial distribution $\bin(N_o,p)$ thus,
from (\ref{eq:binomial}),
\bea
\lefteqn{
\prob^{N_o}\{ S \leq z  | V(\theta^*_k) = p\} =
\sum_{i=0}^{\lfloor z\rfloor} \binom{N_o}{i}p^i(1-p)^{N_o-i}  } \nonumber \rule{2cm}{0cm}
\\
&&= 
\FB(N_o-\lfloor z\rfloor, \lfloor z\rfloor +1,1-p) \nonumber \\
&& = 1-\FB(\lfloor z\rfloor +1,N_o-\lfloor z\rfloor,p) \nonumber\\
&& \leq  1-\FB(z+1,N_o-z,p). \label{eq:cprob_S}
\eea
Considering eq.\ (\ref{eq:scenariobound}), we next let
\beq
F_V(t) \doteq \FB(n,N+1-n ; t) + \Psi(t), \quad t\in[0,1],
\label{eq:scenariobound_eq}
\eeq
where $\Psi(t)$ is some unknown function such that $0\leq \Psi(t)\leq 1-\FB(n,N+1-n ; t)$,  for all $t\in[0,1]$,
and $\Psi(0) = \Psi(1)=0$.
Observe that $\Psi(t)$ is identically zero if the scenario problem is f.s.\ w.p.\  one.
Consider  the event
\beas
\mathrm{GoodTrue} &\doteq &  \{ \mathrm{True} \cap V(\theta_k^*) \leq \epsilon   \} \\
&=&
 \{S \leq \lfloor \epsilon' N_o\rfloor  \cap V(\theta_k^*) \leq \epsilon   \}.
\eeas
Leting $z\doteq \lfloor\epsilon' N_o\rfloor$, we have that
\bea
 \prob^{N+N_o} \{\mathrm{GoodTrue}\} =  \prob^{N+N_o} \{ S \leq z  \cap V(\theta_k^*) \leq \epsilon   \}  \nonumber \\
=  
 \int_{0}^{\epsilon}  \prob^{N_o} \{S \leq z  | V(\theta_k^*) = t   \} \d F_V(t) \nonumber \\
\mbox{[using (\ref{eq:cprob_S})]}  = 
 \int_{0}^{\epsilon} \FB(N_o-z,z+1;1-t)  \d F_V(t) \nonumber \\
\mbox{[using (\ref{eq:scenariobound_eq})]} = 
(1-H_{\epsilon,\epsilon'}(N,N_o)) +
R(\epsilon), \label{eq:prob_Goodtrue_1} \rule{1cm}{0cm}
\eea
where we defined 
\bea
H_{\epsilon,\epsilon'}(N,N_o) &\doteq & 
1-\int_{0}^{\epsilon} \FB(N_o-z,z+1;1-t)\cdot \nonumber \\ && \cdot \mbox{beta}(n,N+1-n ; t) \d t \label{eq:Hdef}\\
R(\epsilon) &\doteq &  \int_{0}^\epsilon  \FB(N_o-z,z+1 ; 1-t)  \d \Psi(t).
\label{eq:Rdef} \rule{1cm}{0cm}
\eea
We next analyze the above two terms.
For the first term, we have
\bea
\lefteqn{
1-H_{\epsilon,\epsilon'}(N,N_o) }  \label{eq:betabinomint} \\
 && = \int_{0}^{\epsilon} \FB(N_o-z,z+1;1-t)  \mbox{beta}(n,N+1-n ; t) \d t
 \nonumber \\
&& \mbox{[using (\ref{eq:cprob_S})]} =
\sum_{i=0}^{z} \int_{0}^{\epsilon} 
 \binom{N_o}{i}t^i(1-t)^{N_o-i}
 \cdot \nonumber \\ && \rule{3cm}{0cm}\cdot \mbox{beta}(n,N+1-n ; t) \d t    \nonumber \\
&& = \sum_{i=0}^{z} \int_{0}^{\epsilon} 
 \binom{N_o}{i} \frac{1}{B(n,N+1-n)} t^{i+n-1}(1-t)^{N_o-i+N-n} \d t \nonumber \\
&& = \sum_{i=0}^{z} \int_{0}^{\epsilon}  \binom{N_o}{i} \frac{B(i+n,N_o-i+N-n+1)}{B(n,N+1-n)}
\cdot\nonumber \\ &&  \rule{3cm}{0cm} \cdot\mbox{beta}  (i+n,N_o-i+N-n+1;t)  \d t \nonumber \\
&& \mbox{[by def.\ in (\ref{eq:betabinomial})]}
= \sum_{i=0}^{z} \int_{0}^{\epsilon} \fbb(N_o, n,N+1-n; i) \cdot \nonumber\\ &&  \rule{3cm}{0cm}\cdot \mbox{beta}  (i+n,N_o-i+N-n+1;t)  \d t
\nonumber 		\\
&&= \sum_{i=0}^{z} \fbb(N_o, n,N+1-n; i) 	\cdot \nonumber\\  && \rule{3cm}{0cm}\cdot\FB  (n+i,N+N_o-n-i+1;\epsilon).  
\nonumber
\eea
Observe that, for all $i=0,\ldots,z$, it holds that
\beas
\lefteqn{
\FB(n+i, N+N_o-n-i+1;\epsilon)  } \\
&& = \sum_{j=n+i}^{N+N_o} \binom{N+N_o}{j}\epsilon^j(1-\epsilon)^{N+N_o-j} \\
&& \geq   \sum_{j=n+z}^{N+N_o}\binom{N+N_o}{j}\epsilon^j(1-\epsilon)^{N+N_o-j} \\
&& =  \FB(n+z, N+N_o-n-z+1;\epsilon) .
\eeas
Therefore, we obtain following  bound 
\beas
\lefteqn{
1-H_{\epsilon,\epsilon'}(N,N_o) } \\
& &\geq  \FB(n+z, N+N_o-n-z+1;\epsilon)  \\ && \rule{3cm}{0cm} \cdot\sum_{i=0}^{z} \fbb(N_o, n,N+1-n; i) \nonumber \\
&& = \left(1-\FB(N+N_o-n-z+1,n+z;1-\epsilon) \right)\cdot
 \\ && \rule{3cm}{0cm} \cdot \left(1-H_{1,\epsilon'}(N,N_o) \right).
\label{eq:1_H_bound}
\eeas
For $z = \lfloor\epsilon' N_o\rfloor$, we have, in particular, that
{
\bea
\lefteqn{
1-H_{\epsilon,\epsilon'}(N,N_o) } \nonumber \\
&&\geq  \left(1-\FB(N+N_o-n-\lfloor\epsilon' N_o\rfloor+1,n+\lfloor\epsilon' N_o\rfloor;1-\epsilon) \right)  \nonumber \\   && \rule{2cm}{0cm} \cdot \left(1-H_{1,\epsilon'}(N,N_o) \right) \nonumber \\
&&\geq  \left(1-\FB(N+(1-\epsilon')N_o-n+1,n+\epsilon' N_o;1-\epsilon) \right) \nonumber 
\\   && \rule{2cm}{0cm} \cdot \left(1-H_{1,\epsilon'}(N,N_o) \right) \nonumber \\
& & \doteq  \left(1- \bar\beta_{\epsilon,\epsilon'}(N,N_o) \right)\cdot \left(1-H_{1,\epsilon'}(N,N_o) \right).
\label{eq:1_H_bound2}
\eea
}
We next consider the $R(\epsilon)$ term in (\ref{eq:Rdef}). We have that
\bea
\lefteqn{
R(\epsilon)  =    \int_{0}^\epsilon  (1-\FB(z+1,N_o-z; t) ) \d \Psi(t) \label{eq:Rdef_1} } \\
&& =  \Psi(\epsilon) - \int_{0}^\epsilon \FB(z+1,N_o-z; t)  \d \Psi(t) \nonumber \\
&& \mbox{[integrating by parts]}  = 
\Psi(\epsilon) \FB(N_o-z,z+1; 1-\epsilon) \nonumber \\
&& \rule{2cm}{0cm}+
    \int_{0}^\epsilon \Psi(t)\mbox{beta}(z+1,N_o-z; t)  \d t .  \nonumber
\eea
Since $\Psi(t)\geq 0$ forall $t\in[0,1]$, the above expression shows that $R(\epsilon)\geq 0$
for all $\epsilon\in[0,1]$, with $R(\epsilon)$ being identically zero for problems that are
f.s.
Considering (\ref{eq:prob_Goodtrue_1}), this fact permits us to conclude that
\bea
 \prob^{N+N_o} \{\mathrm{GoodTrue}\} &=&
 \prob^{N+N_o} \{ S \leq z  \cap V(\theta_k^*) \leq \epsilon   \}  \nonumber
\\ & =  & 1-H_{\epsilon,\epsilon'}(N,N_o) +R(\epsilon) \nonumber \\
&\geq & 1-H_{\epsilon,\epsilon'}(N,N_o) \nonumber \\
\mbox{[from (\ref{eq:1_H_bound2})]} &\geq & \left(1- \bar\beta_\epsilon(N,N_o) \right)\cdot\nonumber \\
&& \cdot \left(1-H_{1,\epsilon'}(N,N_o) \right), \nonumber
\eea
which proves (\ref{eq:goodtruebound_exact}) and (\ref{eq:goodtruebound}).
Also, we obtain that
\beas
\prob^{N+N_o} \{\mathrm{True}\} & \doteq &
 \prob^{N+N_o} \{ S \leq z  \}  \nonumber \\
&=& \prob^{N+N_o} \{ S \leq z  \cap V(\theta_k^*) \leq 1   \}  \nonumber \\
&=& 1-H_{1,\epsilon'}(N,N_o) + R(1) \\
&\geq & 1-H_{1,\epsilon'}(N,N_o),
\eeas
which proves (\ref{truebound}).
Further, using (\ref{eq:prob_Goodtrue_1}), we have that
\beas
\prob^{N+N_o} \{\mathrm{BadTrue}\} & \doteq &
 \prob^{N+N_o} \{ S \leq z  \cap V(\theta_k^*) > \epsilon   \}  \nonumber \\
&=& \prob^{N+N_o} \{\mathrm{True}\} \\ &&  - \prob^{N+N_o} \{\mathrm{GoodTrue}\}  \\
&=& H_{\epsilon,\epsilon'}(N,N_o)- H_{1,\epsilon'}(N,N_o) \\ && + (R(1) -R(\epsilon)).
\eeas
If the scenario problem is f.s., then $R(1)=R(\epsilon)=0$, hence
\beas
\prob^{N+N_o} \{\mathrm{BadTrue}\} & =  & H_{\epsilon,\epsilon'}(N,N_o)- H_{1,\epsilon'}(N,N_o) \\
&&  + (R(1) -R(\epsilon)) \\
\mbox{[if problem is f.\ s.]} &=&  H_{\epsilon,\epsilon'}(N,N_o)- H_{1,\epsilon'}(N,N_o)  \\
\mbox{[using (\ref{eq:1_H_bound2})]} &\leq & \bar\beta_{\epsilon,\epsilon'}(N,N_o)(1-H_{1,\epsilon'}(N,N_o)).
\eeas
All the above proves (\ref{eq:badtruebound_exact}).
To upper bound the probability of $\mathrm{BadTrue}$ in the non-fully supported case, we reason instead as follows:
\beas
\lefteqn{
\prob^{N+N_o} \{\mathrm{BadTrue}\} } \\
&& = 
 \int_{\epsilon}^1  \prob^{N_o} \{ S \leq z  | V(\theta_k^*) = t   \} \d F_V(t) \\
&& =  \int_{\epsilon}^1  \FB(N_o-z,z+1;1-t) \d F_V(t)  \\
&& \mbox{[integrand is decreasing in $t$]} \\
&& \leq  \int_{\epsilon}^1  \FB(N_o-z,z+1;1-\epsilon) \d F_V(t) \\
&& = \FB(N_o-z,z+1;1-\epsilon) \int_{\epsilon}^1 \d F_V(t) \\
&& = \FB(N_o-z,z+1;1-\epsilon) \cdot   \prob^{N}\{V(\theta_k^*) > \epsilon \}  \\
&& \mbox{[from (\ref{eq:scenariobound})]} \\
&& \leq  
\FB(N_o-z,z+1;1-\epsilon) \cdot
\FB(N+1-n,n;1-\epsilon) \\
&& \mbox{[since $z=\lfloor \epsilon' N_o \rfloor$]} \\
&& \leq 
\FB((1-\epsilon')N_o, \epsilon' N_o+1;1-\epsilon) \cdot \beta_\epsilon (N),
\eeas
which proves (\ref{eq:badtruebound_exact}).
\qed

\subsection{Proof of  Theorem~\ref{thm:RSD_RVO}}
\label{app:thm:main}
Let us define the event $\mathrm{BadExit}_k$ as the one where the algorithm reaches the
$k$-th iteration, and then exits with a ``bad'' solution, i.e., with a solution $\theta^*_k$ for which $V(\theta^*_k) > \epsilon$. The probability of this event
is the probability that the $\epsilon'$-RVO returns {\tt false} precisely $k-1$ times
(for this guarantees that we reach the $k$-th iteration), and then the event $\mathrm{BadTrue}$
happens at the $k$-th iteration. Therefore, letting $q$ denote the probability of $\mathrm{BadTrue}$,
and $p$ denote the probability of $\mathrm{True}$ (events defined as in Lemma~\ref{lem:main})
we have that
\beas
\prob^{\times\times}\{ \mathrm{BadExit}_k\} &=& (1-p)^{k-1} q .
\eeas
The event $\mathrm{BadExit}$ in which the algorithm terminates with a bad solution is the union of the non-overlapping events
$\mathrm{BadExit}_k$, $k=1,2,\ldots$, therefore
\beas
\prob^{\times\times}\{ \mathrm{BadExit}\} &=& \sum_{k=1}^\infty \prob^{\times\times}\{ \mathrm{BadExit}_k\} \\
&=& \sum_{k=1}^\infty (1-p)^{k-1} q = q \sum_{k=0}^\infty (1-p)^{k} \\
&=& \frac{q}{p} = \frac{\prob^{N+N_o}\{ \mathrm{BadTrue}\}}{\prob^{N+N_o}\{ \mathrm{True}\}}.
\eeas
We now use  (\ref{eq:badtruebound}) to upper bound
$q$, and then use (\ref{truebound}) to conclude that
\beas
\lefteqn{
\prob^{\times\times}\{ \mathrm{BadExit}\} } \\
&&  \leq 
\frac{\FB((1-\epsilon')N_o, \epsilon' N_o+1;1-\epsilon) \beta_\epsilon (N)}{1-H_{1,\epsilon'}(N,N_o)},
\eeas
which proves (\ref{eq:badexitbound}).
In the fully supported case, we can instead use (\ref{eq:badtruebound_exact}) to upper bound
$q$, and hence conclude that
\[
\prob^{\times\times}\{ \mathrm{BadExit}\} \leq \bar\beta_{\epsilon,\epsilon'}(N,N_o),
\]
which proves (\ref{eq:badexitbound_exact}).

Let next $K$ denote the running time of Algorithm~\ref{alg:RSD_RVO}, that is 
the value of the iteration count when the algorithm terminates.
Since the algorithm terminates as soon as a $\mathrm{True}$ 
event
happens, and since the $\mathrm{True}$ 
events are statistically independent among iterations,
 we have that 
$\{K = k\}$ has geometric probability
$(1-p)^{k-1}p$, where $p$ is the probability of $\mathrm{True}$.
Therefore, the expected value of $K$ is $1/p\leq 1/(1-H_{1,\epsilon'}(N,N_o))$,
where 
the  inequality follows from (\ref{truebound}), and this proves point 2 in the theorem.
Via the same reasoning, 
$\{ K > k \}$ has probability
$(1-p)^k$, and hence  we conclude that
\[
\prob^{\times\times}\{K \leq k \} = 1- (1-p)^k \geq  1- H_{1,\epsilon'}(N,N_o)^k,
\]
which proves the third point in
the theorem. \qed

\subsection{Proof of  Corollary~\ref{cor:approxbound}}
\label{app:cor:approxbound}
From from eq.\ (\ref{eq:betabinomint}) we have that, for 
$z\doteq
 \lfloor \epsilon' N_o \rfloor$,
\bea
\lefteqn{
1-H_{1,\epsilon'}(N,N_o)  = }\label{eq:H1_bound1}
\\&&   \int_{0}^{1} \FB(N_o-z,z+1;1-t)  \mbox{beta}(n,N+1-n ; t) \d t.
\nonumber
\eea
We recall that a $\mbox{beta}(\alpha,\beta)$ density has mean $\alpha/(\alpha+\beta)$, peak (mode) at
$(\alpha-1)/(\alpha+\beta-2)$, and variance $\sigma^2 = \alpha\beta/((\alpha+\beta)^2(\alpha+\beta+1))$.
Then, we observe that 
$\FB(N_o-z,z+1;1-t) = 1 - \FB( z+1, N_o-z;t)$, where $\FB( z+1, N_o-z;t)$
is the cumulative distribution of a $\mbox{beta}( z+1, N_o-z)$ density.
The peak of this density is at
$z/(N_o-1)$, which tends to $\epsilon'$ for $N_o\to \infty$; further,
 the variance of this distribution goes to zero
as $O(N_o\inv)$, which permits us to argue that, for large $N_o$, the function $\FB(N_o-z,z+1;1-t)$
has an inflection point near $\epsilon'$ and decreases rapidly from value $\simeq 1$ to
value $\simeq 0$ as $t$ crosses $\epsilon'$. That is, as $N_o\to\infty$, the function $\FB(N_o-z,z+1;1-t)$ tends to a step function which is one for $t< \epsilon'$ and zero for $t > \epsilon'$.
Therefore, we have for the integral in (\ref{eq:H1_bound1}) that
\beas
\lefteqn{
1-H_{1,\epsilon'}(N,N_o)  \to    \int_{0}^{\epsilon'} 1 \cdot  \mbox{beta}(n,N+1-n ; t) \d t } \\
&& 
= \FB(n,N+1-n; \epsilon') = 1-\beta_{\epsilon'}(N),
\eeas
which proves (\ref{eq:H1_bound}).
\qed


\end{document}